\documentclass[11pt]{article}
\usepackage{amssymb}
\usepackage{graphicx}
\usepackage{theorem}
\usepackage{times}

\theoremheaderfont{\scshape}
\theoremstyle{break}
\theorembodyfont{\upshape}
\newtheorem{theorem}{Theorem}
\newtheorem{proposition}{Proposition}

\newtheorem{definition}{Definition}

\newcommand{\be}{\begin{equation}}
\newcommand{\ee}{\end{equation}}
\newcommand{\bea}{\begin{eqnarray}}
\newcommand{\eea}{\end{eqnarray}}
\newcommand{\cov}{{\rm Cov}}

\begin{document}

\title{Testing the Gaussian Copula Hypothesis\\ for
Financial Assets Dependences\footnote{We acknowledge helpful discussions and
exchanges with J. Andersen, P. Embrechts, J.P. Laurent, F. Lindskog, V.
Pisarenko and R. Valkanov. This work was partially supported by 
the James S. Mc Donnell Foundation 21st century scientist award/studying
complex system.}}

\author{Y. Malevergne$^{1,2}$ and D. Sornette$^{1,3}$ \\
\\
$^1$ Laboratoire de Physique de la Mati\`ere Condens\'ee CNRS UMR 6622\\
Universit\'e de Nice-Sophia Antipolis, 06108 Nice Cedex 2, France\\
$^2$ Institut de Science Financi\`ere et d'Assurances - Universit\'e Lyon I\\
43, Bd du 11 Novembre 1918, 69622 Villeurbanne Cedex\\
$^3$ Institute of Geophysics and Planetary Physics
and Department of Earth and Space Science\\
University of California, Los Angeles, California 90095, USA\\
\\
Corresponding author: D. Sornette\\
Institute of Geophysics and Planetary Physics\\
University of California, Los Angeles, California 90095, USA\\
email: sornette@ess.ucla.edu~~~~
tel: (310) 825 28 63 ~~~~~Fax: (310) 206 30 51
}
\date{}
\maketitle

\begin{abstract}
Using one of the key property of copulas
that they remain invariant under an arbitrary monotonous change of variable,
we investigate the null hypothesis that
the dependence between financial assets can be modeled
by the Gaussian copula. We find that most pairs of currencies
and pairs of major stocks are compatible with the Gaussian copula hypothesis,
while this hypothesis can be rejected for the dependence between pairs
of commodities (metals).
Notwithstanding the apparent qualification of the Gaussian copula
hypothesis for most of the currencies and the stocks,
a non-Gaussian copula, such as the Student's copula, cannot be
rejected if it has sufficiently
many ``degrees of freedom''. As a consequence, it may be very 
dangerous to embrace blindly the Gaussian copula hypothesis, especially
when the correlation coefficient between the pair of asset is too high
as the tail dependence neglected by the Gaussian copula can be as
large as $0.6$, i.e., three out five extreme events which occur in unison are missed.

\end{abstract}



\section{Introduction}

The determination of the dependence between assets underlies
many financial activities, such as risk assessment and portfolio management, as
well as option pricing and hedging.
Following \cite{Markovitz}, the covariance and correlation matrices
have, for a long time, been considered as the main tools for quantifying the
dependence between assets. But the dimension of risk captured by the
correlation matrices is only satisfying for elliptic distributions and
for moderate risk amplitudes \cite{Sorphysrep}. In all
other cases, this measure of risk is severely incomplete and can lead to a very
strong underestimation of the real incurred risks \cite{EMN99}.

Although the unidimensional (marginal) distributions of asset returns
are reasonably
constrained by empirical data and are more or less satisfactorily
described by a power law with tail index
ranging between 2 and 4 \cite{V94,Lux96,P96,GDDMOP97,GMAS98} or by
stretched exponentials \cite{Lahsor,GJ99,Sorphysrep,Sorandsim}, no
equivalent results have been obtained
for {\it multivariate} distributions of asset returns. Indeed, a
brute force determination of multivariate distributions is unreliable due
to the limited data set (the curse of dimensionality), while the sole
knowledge of marginals (one-point statistics) of each asset
is not sufficient to obtain information on
the multivariate distribution of these assets which involves all the
$n$-points statistics.

Some progress may be expected from the
concept of copulas, recently proposed to be useful for financial applications
\cite{EMS01, FV98, Haas99, KP99}. This concept has the desirable
property of decoupling the study of the marginal distribution
of each asset from the study of their
collective behavior or dependence. Indeed, the dependence
between assets is entirely embedded in the copula, so that a copula
allows for a
simple description of the dependence structure between assets independently
of the marginals. For instance, assets can have power law marginals
and a Gaussian
copula or alternatively Gaussian marginals and a non-Gaussian copula, and any
possible combination thereof. Therefore, the determination of the multivariate
distribution of assets can be performed in two steps~: (i)
an independent determination of the marginal distributions using
standard techniques
for distributions of a single variable~; (ii) a study of the nature of
the copula
characterizing completely the dependence between the assets.
This exact separation between the marginal distributions and the dependence
is potentially very useful for risk management or option pricing and
sensitivity analysis since it allows for testing several scenarios with
different kind of dependences between assets while the marginals can be set to
their well-calibrated empirical estimates.  Such an approach has been used by
\cite{EHJ01} to provide various bounds for the Value-at-Risk of a portfolio
made of depend risks, and by \cite{R99} or \cite{CL2000} to price and to
analyse the pricing sensitivity of binary digital options or options on the
minimum of a basket of assets.

A fundamental limitation of the copula approach is that there is
in principle an infinite number of possible copulas
\cite{GM86,G87,GR93,J93,Nelsen} and, up to now,
no general empirical study has
determined the classes of copulas that are acceptable for financial
problems. In general, the choice of a given copula is guided both by
the empirical
evidences and the technical constraints, i.e., the number of
parameters necessary to
describe the copula, the possibility to obtain efficient
estimators of these parameters and also the possiblity offered by the chosen
parameterization to allow for tractable analytical calculation. It is
indeed sometimes
more advantageous to prefer a simplest copula to one that fit better
the data, provided that we can clearly quantify the effects of this
substitution.

In this vein, the first goal of the present article is to show
that, in most cases, the Gaussian copula can provide an
approximation of the unknown true copula that is sufficiently good so
that it cannot be rejected
on a statistical basis. Our second goal is to draw the
consequences of the parameterization involved in the Gaussian copula
in term of potential over/underestimation of the risks, in particular for large
and extreme events.

The paper is organized as follows.

In section 2, we first recall some important general definitions and
theorems about
copulas that will be useful in the sequel.
We then introduce the concept of tail dependence that will allow
us to quantify the probability that two extreme events might occur
simultaneously.
We define and describe the two copulas that will be at the
core of our study~: the Gaussian copula and the Student's copula and compare
their properties particularly in the tails.

In section 3, we present our statistical testing procedure which is applied
to pairs of financial time series. First of all, we
determine a test statistics which leads us to
compare the empirical distribution of the data with a
$\chi^2$-distribution using a bootstrap method. We also test the
sensitivity of our procedure by applying it to synthetic multivariate Student's
time series. This allows us to determine the minimum statistical test value
needed to be able to distinguish between a Gaussian and a Student's
copula, as a function of the number of degrees of freedom and of the
correlation strength.

Section 4 presents the empirical results obtained for the following assets
which are combined pairwise in the test statistics:
\begin{itemize}
\item 6 currencies,
\item 6 metals traded on the London Metal Exchange,
\item 22 stocks choosen among the largest companies quoted on the New
York Stocks Exchange.
\end{itemize}
We show that the Gaussian copula hypothesis is very reasonnable for
most stocks and currencies, while it is hardly compatible with the
description of multivariate behavior for metals.

Section 5 summarizes our results and concludes.

\section{Generalities about copulas}
\subsection{Definitions and important results about copulas}

This section does not pretend to provide
a rigorous mathematical exposition
of the concept of copula. We only recall a few basic definitions and
theorems that will be useful in the following (for
more information about the concept of copula, see for instance
\cite{Lindskog, Nelsen}).

We first give the definition of a copula of $n$ random variables.
\begin{definition}[Copula]
A function  $C$ : $[0,1]^n \longrightarrow [0,1]$ is a $n$-copula if it
enjoys the following properties~:
\begin{itemize}
\item $\forall u \in [0,1] $, $C(1,\cdots, 1, u, 1 \cdots, 1)=u$~,
\item $\forall u_i \in [0,1] $, $C(u_1, \cdots, u_n)=0$ if
at least one of the $u_i$ equals zero~,
\item $C$ is grounded and $n$-increasing, i.e., the $C$-volume of every
boxes whose vertices lie in $[0,1]^n$ is positive.
\end{itemize}
\end{definition}

It is clear from this definition that
a copula is nothing but a multivariate distribution with support in
[0,1]$^n$ and with uniform marginals. The fact that such copulas
can be very useful for representing
multivariate distributions with arbitrary marginals is seen from the
following result.
\begin{theorem}[Sklar's Theorem]
Given an $n$-dimensional
distribution function  $F$ with {\it continuous} marginal (cumulative) distributions
$F_1, \cdots, F_n$, there
exists a {\it unique} $n$-copula $C$ : $[0,1]^n \longrightarrow [0,1]$ such
that~: \be
F(x_1, \cdots, x_n) = C(F_1(x_1), \cdots, F_n(x_n))~.
\ee
\end{theorem}

This theorem provides both a parameterization of
multivariate distributions and a construction scheme for copulas.
Indeed, given
a multivariate distribution $F$ with marginals $F_1, \cdots, F_n$, the function
\be
\label{eq:cop}
C(u_1, \cdots, u_n)= F\left( F_1^{-1}(u_1), \cdots, F_n^{-1}(u_n) \right)
\ee
is automatically a $n$-copula. This copula is the copula of the
multivariate distribution $F$.
We will use this method in the sequel
to derive the
expressions of standard copulas such as the Gaussian copula or the Student's
copula.

A very powerful property of copulas is their invariance under
arbitrary strictly increasing mapping of the random variables~:
\begin{theorem}[Invariance Theorem]
Consider $n$ {\it continuous} random variables  $X_1, \cdots ,
X_n$ with copula $C$.
Then, if $g_1(X_1), \cdots, g_n(X_n)$ are
strictly increasing on the ranges of $X_1, \cdots , X_n$, the random variables
$Y_1=g_1(X_1), \cdots, Y_n=g_n(X_n)$ have exactly the same copula $C$.
\end{theorem}
It is this result that shows us that the full dependence between the $n$
random variables is completely captured by the copula, independently of the
shape of the marginal distributions. This result is at the basis
of our statistical study presented in section 3.

\subsection{Dependence between random variables}

The dependence between two time series is usually described by their
correlation coefficient. This measure is fully satisfactory only for
elliptic distributions \cite{EMN99}, which are functions of a quadratic form
of the random variables, when one is interested in moderately size events.
However, an important issue for risk
management concerns the determination of the dependence of the distributions
in the tails. Practically, the question is whether it is more probable
that large or extreme events occur simultaneously or on the contrary
more or less independently.
This is refered to as the presence or abscence of ``tail dependence''.

The tail dependence is also an interesting concept in studying 
the {\it contagion} of crises between markets or countries.
These questions have recently been addressed by
\cite{AC01,LS01,S99} among several others.  Large negative
moves in a country or market are often found to imply large negative moves in
others.

Technically, we need to determine the probability
that a random variable $X$ is large, knowing that the random variable $Y$ is
large.
\begin{definition}[Tail dependence 1]
    Let $X$ and $Y$ be random variables with
{\it continuous} marginals $F_X$ and $F_Y$. The (upper) tail dependence
coefficient of $X$ and $Y$ is, if it exists,
\be
\lim_{u \rightarrow 1} \Pr\{ X > F_X^{-1}(u) | Y > F_Y^{-1}(u) \}=
\lambda \in [0,1]~.
\ee
In words, given that $Y$ is very large (which occurs with probability 
$1-u$), the probability that $X$ is very large at the same probability
level $u$ defines asymptotically the tail dependence coefficient $\lambda$.
\end{definition}
It turns out that this
tail dependence is a pure copula property which is independent of the
marginals. Let $C$ be the copula of the variables $X$ and $Y$, then
\begin{theorem}
if the bivariate copula $C$ is such that
\be
\lim_{u \rightarrow 1} \frac{\bar C(u,u)}{1-u}=\lambda
\ee
exists (where $\bar C(u,u) = 1-2u-C(u,u)$),
then $C$ has an upper tail dependence coefficient $\lambda$.
\end{theorem}

If $\lambda > 0$, the copula presents tail dependence and large events tend to
occur simultanously, with the probabilty $\lambda$. On the contrary,
when $\lambda=0$, the copula has no tail dependence in this sense
and large events appear to occur essentially independently.
There is however a subtlety in this definition of tail
dependence. To make it clear, first consider the case where for large 
$X$ and $Y$
the distribution function $F(x,y)$ factorizes such that
\be
\label{eq:tl}
\lim_{x,y \rightarrow \infty} \frac{F(x,y)}{F_X(x)F_Y(y)}=1~.
\ee
This means that, for $X$ and $Y$ sufficiently large, these two
variables can be considered as independent. It is then easy to show that
\bea
\lim_{u \rightarrow 1} \Pr\{ X > F_X^{-1}(u) | Y > F_Y^{-1}(u) \}&=&
\lim_{u \rightarrow 1} 1-F_X( F_X^{-1}(u)) \\
&=&\lim_{u \rightarrow 1} 1-u =0,
\eea
so that independent variables really have no tail dependence, as one
can expect.

Unfortunatly, the converse does not holds~:
a value $\lambda = 0$ does not automatically imply
true independence, namely that $F(x, y)$ satisfies equation (\ref{eq:tl}).
Indeed, the tail independence criterion $\lambda = 0$ may still be
associated with
an absence of factorization of the multivariate distribution for large
$X$ and $Y$. In a weaker sense, there
may still be a dependence in the tail even when $\lambda = 0$.
Such behavior is for instance exhibited by the Gaussian
copula, which has zero tail dependence according to the definition 2
but nevertheless does not have a factorizable multivariate
distribution, since the non-diagonal term of the quadratic form in the
exponential function does not become negligible in general as $X$ and
$Y$ go to infinity. To summarize, the {\it tail independence}, according to
definition 2, is not equivalent to the {\it independence in the tail}
as defined in equation (\ref{eq:tl}).

After this brief review of the main concepts underlying copulas, we now present
two special families of copulas~: the Gaussian copula and the Student's copula.

\subsection{The Gaussian copula}

The Gaussian copula is the copula derived from the multivariate
Gaussian distribution. Let $\Phi$ denote the standard Normal (cumulative) distribution and
$\Phi_{\rho,n}$ the $n$-dimensional Gaussian distribution with correlation
matrix ${\bf \rho}$.  Then, the Gaussian $n$-copula
with correlation matrix ${\bf \rho}$ is \be
\label{eq:gc}
C_\rho(u_1, \cdots, u_n) = \Phi_{\rho,n} \left(\Phi^{-1}(u_1),
\cdots, \Phi^{-1}(u_n) \right)~,
\ee
whose density
\be
c_\rho(u_1, \cdots, u_n) = \frac{\partial C_\rho(u_1, \cdots,
u_n)}{\partial u_1 \cdots \partial u_n}
\ee
reads
\be
\label{eq:gcd}
c_\rho(u_1, \cdots, u_n) = \frac{1}{\sqrt{\det {\bf \rho}}} \exp \left( -
\frac{1}{2} y_{(u)}^t
({\bf \rho}^{-1}- \mbox{Id}) y_{(u)} \right)
\ee
with $y_k(u)=\Phi^{-1}(u_k)$.  Note that theorem 1 and equation
(\ref{eq:cop}) ensure that $C_\rho(u_1, \cdots, u_n)$ in equation (\ref{eq:gc})
is a copula.

As we said before, the Gaussian copula does not have a tail dependence~:
\be
\lim_{u \rightarrow 1} \frac{\bar C_\rho(u,u)}{1-u} = 0,~ ~ \forall
\rho \in (-1, 1).
\ee
This results is derived for example in \cite{EMS01}. But this does not mean
that the Gaussian copula goes to the independent (or product) copula
$\Pi(u_1,u_2) = u_1 \cdot u_2$ when $(u_1, u_2)$ goes to one. Indeed, consider
a distribution $F(x,y)$ with Gaussian copula :
\be
F(x,y) = C_\rho(F_X(x), F_Y(y)).
\ee
Its density is
\be
f(x,y)= c_\rho(F_X(x), F_Y(y)) \cdot f_X(x) \cdot f_Y(y),
\ee
where $f_X$ and $f_Y$ are the densities of $X$ and $Y$. Thus,
\be
\lim_{(x,y) \rightarrow \infty} \frac{f(x,y)}{f_X(x) \cdot f_Y(y)} = \lim_{(x,y) \rightarrow \infty}
c_\rho(F_X(x), F_Y(y)), \ee
which should equal 1 if the variables $X$ and $Y$ were independent
in the tail. Reasoning in the quantile space, we set $x=F_X^{-1}(u)$ and
$y=F_Y^{-1}(u)$, which yield
\be
\lim_{(x,y) \rightarrow \infty} \frac{f(x,y)}{f_X(x) \cdot f_Y(y)} = \lim_{u
\rightarrow 1} c_\rho(u,u).
\ee

Using equation (\ref{eq:gcd}), it is now obvious to show that $c_\rho(u,u)$
goes to one when $u$ goes to one, if and only if $\rho=0$ which is equivalent to
$C_{\rho=0}(u_1,u_2) = \Pi(u_1,u_2)$ for every $(u_1,u_2)$. When $\rho>0$, 
$c_\rho(u,u)$ goes to infinity, while for $\rho$ negative,
$c_\rho(u,u)$ goes to zero as $u \to 1$. Thus, the dependence structure described by the
Gaussian copula is very different from the dependence structure of the
independent copula, except for $\rho=0$.

The Gaussian copula is completly determined by the knowledge
of the correlation matrix ${\bf \rho}$. The parameters involved in the
description of the Gaussian copula are very simple to estimate, as we
shall see in the following.

In our tests presented below, we focus on pairs of assets, i.e., on Gaussian
copulas involving only two random variables. Testing the Gaussian copula
hypothesis for two random variables gives useful information for a larger
number of dependent variables constituting a large basket or
portfolio. Indeed, let us assume that each
pair $(a,b), (b,c)$ and $(c,a)$ have a gaussian copula. Then, the triplet
$(a,b,c)$ has also a Gaussian copula. This result generalizes to an arbitrary
number of random variables.

\subsection{The Student's copula}
\label{sec:stud_cop}

The Student's copula is derived from the Student's multivariate distribution.
Given a multivariate Student's distribution
$T_{\rho,\nu}$ with $\nu$ degrees of freedom and a correlation matrix
${\bf \rho}$
\be
T_{\rho,\nu}( {\bf x} ) =  \frac{1}{\sqrt{\det {\bf \rho}}}
\frac{\Gamma \left( \frac{\nu+n}{2} \right)
}{\Gamma \left(
\frac{\nu}{2} \right) (\pi \nu)^{N/2}}  \int_{-\infty}^{x_1} \cdots \int_{-\infty}^{x_N}
\frac{d{\bf x}}{\left( 1+ \frac{x^t {\bf \rho} x}{\nu} \right)
^{ \frac{\nu +n}{2}}}~,
\ee
 the corresponding Student's copula reads~:
\be
C_{\rho, \nu}(u_1, \cdots, u_n) = T_{\rho, \nu} \left(t_\nu^{-1}(u_1),
\cdots, t_\nu^{-1}(u_n) \right),
\ee
where $t_\nu$ is the univariate Student's distribution with $\nu$ degrees of
freedom. The density of the Student's copula is thus
\be
c_{\rho, \nu}(u_1, \cdots, u_n) =  \frac{1}{\sqrt{\det {\bf \rho}}}
\frac{\Gamma \left( \frac{\nu+n}{2} \right)
\left[ \Gamma \left( \frac{\nu}{2} \right) \right]^{n-1}}{\left[\Gamma \left(
\frac{\nu+1}{2} \right) \right]^n}
\frac{\prod_{k=1}^n \left(1+\frac{y_k^2}{\nu} \right)^{ \frac{\nu
+1}{2}}}{\left( 1+ \frac{y^t {\bf \rho} y}{\nu} \right)
^{ \frac{\nu +n}{2}}}~,
\ee
where $y_k = t_{\nu}^{-1}(u_k)$.

\begin{figure}
\begin{center}
\includegraphics[width=15cm]{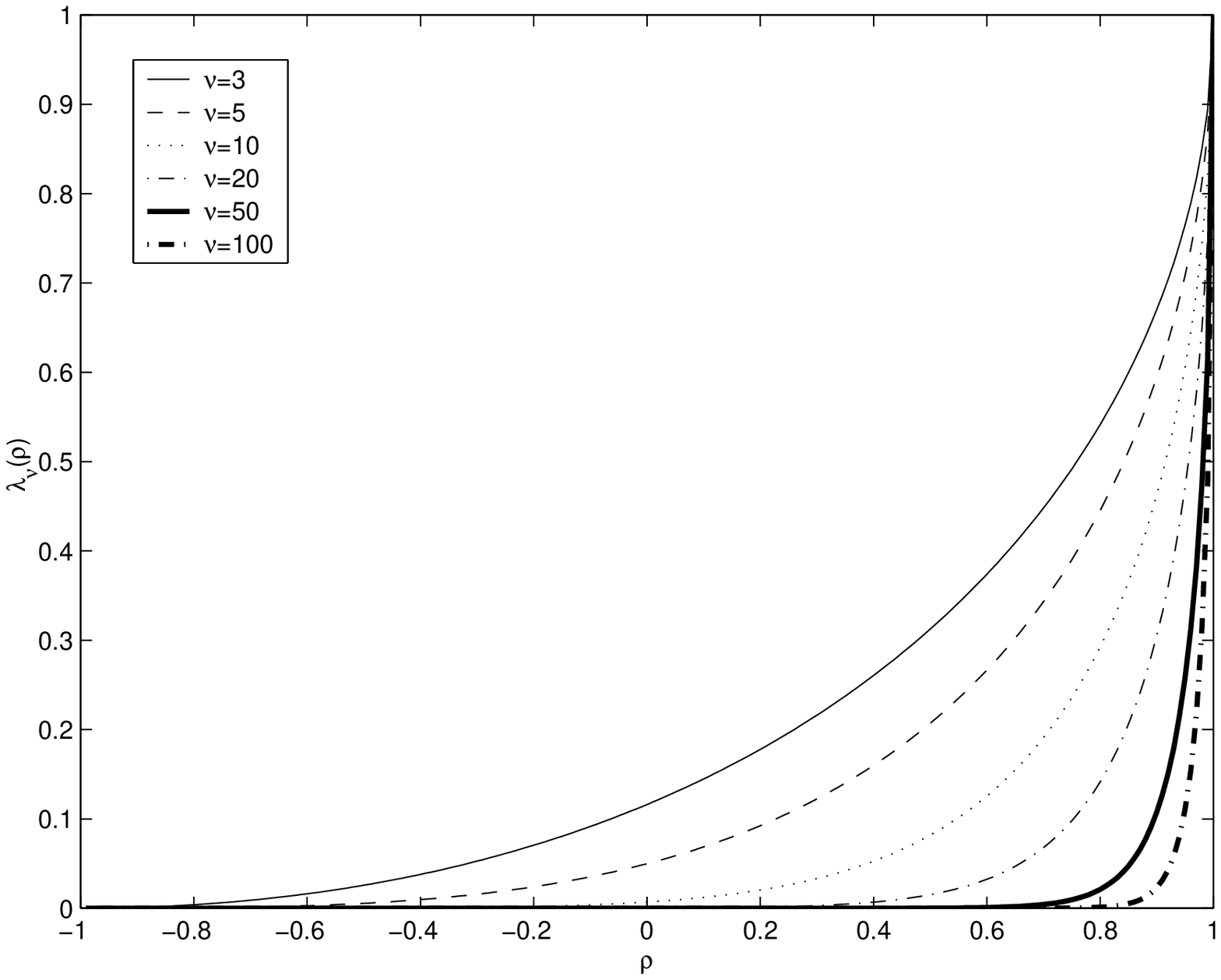}
\caption{\label{fig:1} Upper tail dependence coefficient $\lambda_\nu(\rho)$
for the Student's copula with $\nu$ degrees of freedom as a function of the
correlation coefficient $\rho$, for different values of $\nu$.}
\end{center}
\end{figure}

Since the Student's distribution tends to the normal distribution
when $\nu$ goes to
infinity, the Student's copula tends to the Gaussian copula as $\nu
\to +\infty$.
In contrast to the Gaussian copula, the Student's copula for $\nu$ finite
presents a tail dependence given by~:
\be
\lambda_\nu(\rho)=\lim_{u \rightarrow 1}
\frac{\bar C_{\rho,\nu}(u,u)}{1-u} =2 \bar t_{\nu+1} \left( \frac{\sqrt{\nu+1}
\sqrt{1-\rho}}{\sqrt{1+\rho}} \right)~,
\ee
where $\bar t_{\nu+1}$ is the complementary cumulative univariate Student's
distribution with $\nu+1$ degrees of freedom (see \cite{EMS01} for the proof).
Figure \ref{fig:1} shows the upper tail dependence coefficient as a
function of the correlation coefficient
$\rho$ for different values of the number $\nu$ of degrees of freedom.
As expected from the fact that
the Student's copula becomes identical to the Gaussian copula for
$\nu \to +\infty$ for all $\rho \neq 1$, $\lambda_\nu(\rho)$ exhibits
a regular decay
to zero as $\nu$ increases.
Moreover, for $\nu$ sufficiently large, the tail dependence is significantly
different from $0$ only when the correlation coefficient is
sufficiently close to $1$. This suggests that, for moderate values of the
correlation coefficient, a Student's copula with a large number of
degrees of freedom
may be difficult to distinguish from the Gaussian copula from a
statistical point of
view. This statement will be made quantitative in the following.

\begin{figure}
\begin{center}
\includegraphics[width=15cm]{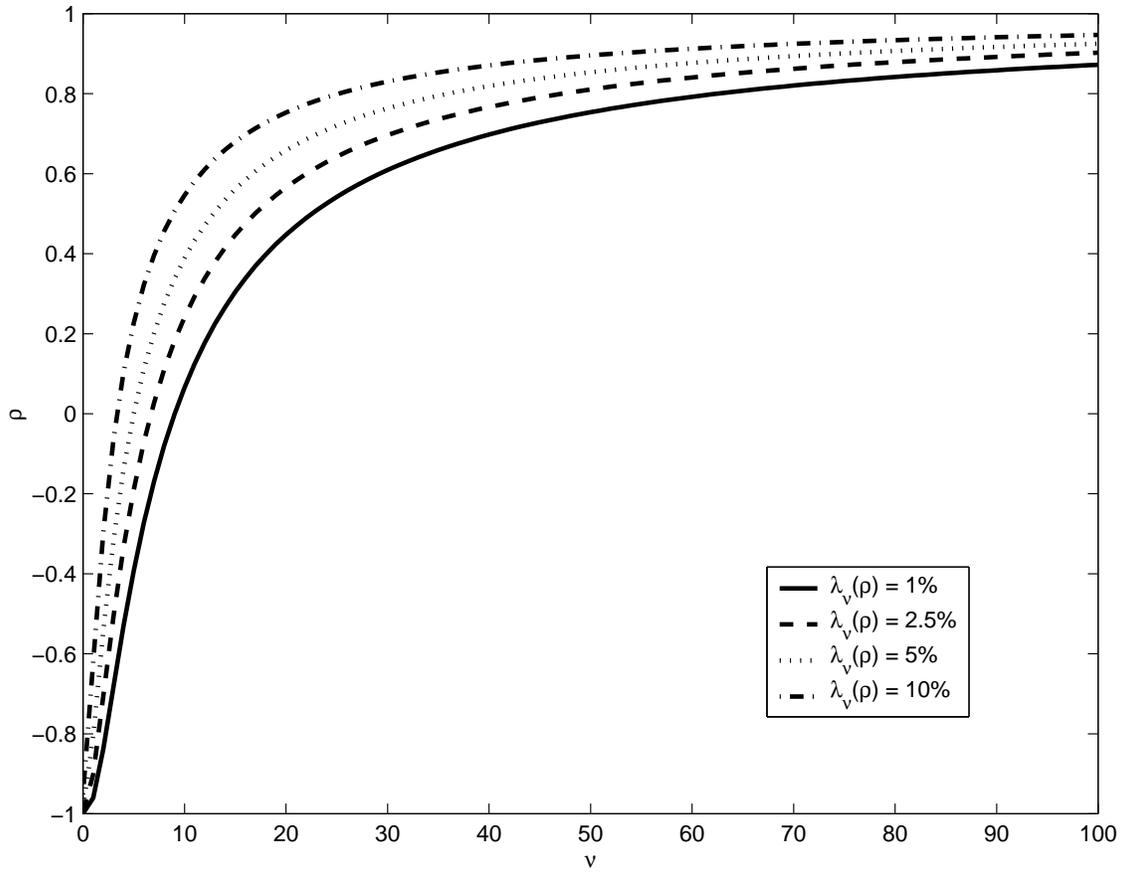}
\caption{\label{fig:tdl} Maximum
value of the correlation coefficient $\rho$ as a function of $\nu$, below
which the tail dependence $\lambda_\nu(\rho)$ of
a Student' copula is smaller than a given
small value, here taken equal to $\lambda_\nu(\rho) =1\%, 2.5\%, 5\%$ 
and $10\%$.
The choice $\lambda_\nu(\rho) = 5\%$ for instance corresponds to $1$ event
in $20$ for which the pair of variables are asymptotically coupled. At the
$1-\lambda_\nu(\rho)$ probability level, values of $\lambda \leq
\lambda_\nu(\rho) $ are
undistinguishable from $0$, which means that
the Student's copula can be approximated by a
Gaussian copula.}
\end{center}
\end{figure}

Figure \ref{fig:tdl} presents the same information in a different way
by showing the maximum
value of the correlation coefficient $\rho$ as a function of $\nu$, below
which the tail dependence $\lambda_\nu(\rho)$ of
a Student's copula is smaller than a given
small value, here taken equal to $1\%, 2.5\%, 5\%$ and $10\%$.
The choice $\lambda_\nu(\rho) = 5\%$ for instance corresponds to $1$ event
in $20$ for which the pair of variables are asymptotically coupled. At the
$95\%$ probability level, values of $\lambda_\nu(\rho) \leq 5\%$ are
undistinguishable from $0$, which means that the Student's copula can
be approximated by a
Gaussian copula.

The description of a Student's copula relies on two
parameters : the correlation matrix ${\bf \rho}$, as in the Gaussian case, and
in addition the number of degrees of freedom $\nu$. The estimation of the
parameter $\nu$ is rather difficult and this has an important impact on the
estimated value of the correlation matrix. As a consequence, the
Student's copula is more
difficult to calibrate and use than the Gaussian copula.

\section{Testing the Gaussian copula hypothesis}

In view of the central role that the Gaussian paradigm has played and
still plays in
particular
in finance, it is natural to start with the simplest choice of
dependence between
different random variables, namely the Gaussian copula. It is also a natural
first step as the Gaussian copula imposes itself in an approach which consists
in (1) performing a nonlinear transformation on the random variables
into Normal
random variables (for the marginals) which is always possible and (2)
invoking a maximum entropy principle (which amounts to add the least
additional information
in the Shannon sense) to construct the multivariable distribution of
these Gaussianized
random variables \cite{Sorphysrep,Sorandsim,AndersenSor}.

In the sequel, we will denote by $H_0$ the null hypothesis according to which
the dependence between two (or more) random variables $X$ and $Y$ can be described
by the Gaussian copula.

\subsection{Test Statistics \label{sectonrhodef}}
We now derive the test statistics which will allow us to reject or not our
null hypothesis $H_0$ and state the following proposition:

\begin{proposition}
Assuming that the $N$-dimensionnal random vector ${\bf x}=(x_1,
\cdots, x_N)$ with distribution function $F$ and marginals $F_i$, satisfies the
null hypothesis $H_0$, then, the variable
\be
z^2 = \sum_{j,i=1}^N \Phi^{-1}(F_i(x_i))~ ({\bf \rho}^{-1})_{ij}~
\Phi^{-1}(F_j(x_j)),
\ee
where the matrix ${\bf \rho}$ is
\be
{\bf \rho}_{ij} = \cov[ \Phi^{-1}(F_i(x_i)), \Phi^{-1}(F_j(x_j))],
\ee
follows a $\chi^2$-distribution with $N$ degrees of freedom.
\end{proposition}

To proove the proposition above, first consider an $N$-dimensionnal random
vector ${\bf x}=(x_1, \cdots, x_N)$. Let us denote by $F$ its distribution
function and by $F_i$ the marginal distribution of each $x_i$. Let us now
assume that the distribution function $F$ satisfies $H_0$, so that $F$ has a
Gaussian copula with correlation matrix ${\bf \rho}$ while the $F_i$'s can be
any distribution function. According to theorem 1, the distribution $F$ can be
represented as~: \be
F(x_1, \cdots, x_N)= \Phi_{{\bf \rho},N} (\Phi^{-1}(F_1(x_1)), \cdots,
\Phi^{-1}(F_N(x_N)))~.
\ee
Let us now transform the $x_i$'s into Normal random variables $y_i$'s~:
\be
y_i = \Phi^{-1}(F_i(x_i))~.   \label{nfmlqa}
\ee
Since the mapping $ \Phi^{-1}(F_i(\cdot))$ is obviously increasing, theorem
2 allows us to conclude that the copula of the variables $y_i$'s is identical
to the copula of the variables $x_i$'s. Therefore, the variables $y_i$'s have
Normal marginal distributions and a Gaussian copula with correlation matrix
${\bf \rho}$. Thus, by definition, the multivariate distribution of the $y_i$'s
is the multivariate Gaussian distribution with correlation matrix ${\bf
\rho}$~: \bea
G({\bf y})&=& \Phi_{{\bf \rho},N} ( \Phi^{-1}(F_1(x_1)), \cdots,
\Phi^{-1}(F_N(x_N)))\label{eq:24} \\
&=& \Phi_{{\bf \rho},N} (y_1, \cdots, y_N),  \label{eq:25}
\eea
and ${\bf y}$ is a Gaussian random vector. From equations
(\ref{eq:24}-\ref{eq:25}), we obviously have
\be
{\bf \rho}_{ij} = \cov[ \Phi^{-1}(F_i(x_i)), \Phi^{-1}(F_j(x_j))].
\ee

Consider now the random variable
\be
\label{eq:z}
z^2 = {\bf y^t \rho^{-1} y} = \sum_{i,j=1}^N y_i~(\rho^{-1})_{ij}~y_j~,
\ee
where $\cdot^t$ denotes the transpose operator.
This variable has already been considered in \cite{Sorphysrep} in
preliminary statistical tests of the transformation (\ref{nfmlqa}).
It is well-known that the variable $z^2$ follows a
$\chi^2$-distribution with $N$ degrees of freedom. Indeed, since
 {\bf y} is a Gaussian random vector with covariance matrix\footnote{Up to now,
the matrix ${\bf \rho}$ was named {\it correlation matrix}. But in fact, since
the variables $y_i$'s have unit variance, their correlation matrix is also
their {\it covariance matrix}.} ${\bf \rho}$, it follows that the
components of the vector
\be
\tilde {\bf y} = {\bf \rho}^{-1/2} {\bf y},
\ee
are {\it independent} Normal random variables. Here, ${\bf \rho}^{-1/2}$
denotes the square root of the matrix ${\bf \rho}^{-1}$, which can be obtain by
the Cholevsky decomposition, for instance. Thus, the sum ${\bf {\tilde
y}^t \tilde y} = z^2$ is the sum of the squares of $N$ independent Normal
random variables, which follows a $\chi^2$-distribution with $N$ degrees of
freedom.

\subsection{Testing procedure}

The testing procedure used in the sequel is now described.
We consider two financial series ($N=2$) of size $T$: $\{x_1(1), \cdots,
x_1(t), \cdots, x_1(T)\}$
and $\{x_2(1), \cdots, x_2(t), \cdots, x_2(T)\}$. We assume that the
vectors ${\bf x}(t)=(x_1(t),x_2(t))$, $t\in \{1,\cdots,T\}$ are
independent and identicaly distributed with distribution $F$, which
implies that the variables $x_1(t)$ (respectively $x_2(t)$), $t\in
\{1,\cdots,T\}$, are also independent and identicaly
distributed, with distributions $F_1$ (respectively $F_2$).

The cumulative distribution $\hat F_i$ of
each variable $x_i$, which is estimated empirically, is given by
\be
\hat F_i(x_i) = \frac{1}{T} \sum_{k=1}^T {\bf 1}_{\{x_i < x_i(k) \}},
\ee
where ${\bf 1}_{\{\cdot \}}$ is the indicator function, which equals
one if its argument is true and zero otherwise. We use these
estimated cumulative distributions to obtain the Gaussian variables
$\hat y_i$ as~:
\be
\hat y_i(k)= \Phi^{-1}\left ( \hat F_i(x_i(k)) \right)~~~~k \in \{1,
\cdots, T\}~.   \label{nglnlkda}
\ee

The sample covariance matrix ${\bf \hat \rho}$ is estimated by the expression~:
\be
\label{eq:rho}
{\bf \hat \rho} = \frac{1}{T} \sum_{i=1}^T {\bf \hat y}(i) \cdot {\bf
    \hat y}(i)^t
\ee
which allows us to calculate the variable
\be
\hat z^2(k) = \sum_{i,j=1}^2 \hat y_i(k)~(\rho^{-1})_{ij}~ \hat y_j(k)~,
\ee
as defined in (\ref{eq:z})
for $k \in \{1, \cdots, T\}$, which
should be distributed according to a $\chi^2$-distribution if the Gaussian
copula hypothesis is correct.

The usual way for comparing an empirical with a theoretical
distribution is to measure the distance between these two
distributions and to perform the Kolmogorov test
or the Anderson-Darling \cite{Anderdarling}
test (for a better accuracy in the tails of the distribution).
The Kolmogorov distance is the maximum local
distance along the quantile which most often occur in the bulk of the 
distribution,
while the Anderson-Darling distance
puts the emphasis on the tails of the two distributions by a suitable 
normalization.
We propose to complement these two distances by two additional measures which
are defined as averages of the Kolmogorov distance and of the Anderson-Darling
distance respectively:
\bea
{\rm Kolmogorov:}~~~d_1 &=& \max_z | F_{z^2}(z^2) - F_{\chi^2}(z^2) |
\label{eq:d1}\\
{\rm average ~Kolmogorov:}~~~d_2 &=& \int | F_{z^2}(z^2) -
F_{z^2}(z^2) |~dF_{\chi^2}(z^2) \\
{\rm Anderson-Darling:}~~~d_3 &=& \max_z \frac{ | F_{z^2}(z^2) -
F_{\chi^2}(z^2)
    |}{\sqrt{F_{\chi^2}(z^2)[1-F_{\chi^2}(z^2)]}}   \label{d3distance}\\
{\rm average ~Anderson-Darling:}~~~d_4 &=& \int \frac{ | F_{z^2}(z^2)
-
F_{\chi^2}(z^2)|}{\sqrt{F_{\chi^2}(z^2)[1-F_{\chi^2}(z^2)]}}~dF_{\chi^2}(
z^2 ) \label{eq:d4}
\eea
The Kolmogorov distance $d_1$ and its
average $d_2$ are more sensitive to the deviations occurring in the bulk of
the distributions. In contrast, the Anderson-Darling distance
$d_3$ and its average $d_4$ are more accurate in the tails
of the distributions. We present our statistical tests for these four distances
in order to be as complete as possible with respect to the different
sensitivity of the tests.

The distances $d_2$ and $d_4$ are not of common use in statistics, so let us
justify our choice. One usually uses distances similar to  $d_2$ and $d_4$ but
which differ by the square instead of the modulus of  $F_{z^2}(z^2)
-F_{\chi^2}(z^2)$ and lead respectively to the $\omega$-test and the
$\Omega$-test, whose statitics are theoretically known.
The main advantage of the distances $d_2$ and $d_4$ with respect to the more
usual distances $\omega$ and $\Omega$ is that they are simply equal to the average
of $d_1$ and $d_3$. This averaging is very interesting and provides important
information. Indeed, the distances $d_1$ and $d_3$ are mainly controlled by the point that
maximizes the argument within the $\max(\cdot)$ function. They are thus
sensitive to the presence of an outlier. By averaging, $d_2$ and $d_4$ become less
sensitive to outliers, since the weight of such points is only of order $1/T$
(where $T$ is the size of the sample) while it equals one for $d_1$ and $d_3$. Of
course, the distances $\omega$ and $\Omega$ also perform a smoothing since they are
averaged quantities too. But they are the average of the square of $d_1$ and $d_3$
which leads to an undesired overweighting of the largest events. In fact, this
weight function is chosen as a convenient analytical form that allows one to
derive explicitely the theoretical asymptotic statistics
for the $\omega$ and $\Omega$-tests. In contrast, using the modulus of $F_{z^2}(z^2)
-F_{\chi^2}(z^2)$ instead of its square in the expression of $d_2$ and $d_4$, no
theoretical test statistics can be derived analytically. In other words, 
the presence of
the square instead of the modulus of $F_{z^2}(z^2) -F_{\chi^2}(z^2)$ in the
definition of the distances $\omega$ and $\Omega$ is motivated by 
mathematical convenience rather than by statistical pertinence. In sum, the sole
advantage of the standard distances $\omega$ and $\Omega$ with respect to the distances
$d_2$ and $d_4$ introduced here is the theoretical knowledge of their distributions.
However, this advantage disappears in our present case in which the covariance
matrix is not known {\it a priori} and needs to be estimated from the empirical data: 
indeed, the exact
knowledge of all the parameters is necessary in the derivation of the theoretical
statistics of the $\omega$ and $\Omega$-tests (as well as the Kolmogorov test).
Therefore, we cannot directly use the results of these standard statistical
tests. As a remedy, we propose a bootstrap method \cite{Bootstrap}, whose
accuracy is proved by \cite{CL97} to be at least as good as that given by
asymptotic methods used to derive the theoretical distributions. For the
present work, we have determined that the generation of 10,000 synthetic time
series was sufficient to obtain a good approximation of the distribution of
distances described above.
Since a bootstrap method is needed to determine the tests statistics in
every case, it is convenient to choose functional forms different from the
usual ones in the $\omega$ and $\Omega$-tests
 as they provide an improvement with respect to statistical reliability, as
 obtained with the $d_2$ and $d_4$ distances introduced here.

To summarize, our test procedure is as follows.
\begin{enumerate}
\item Given the original time series ${\bf x}(t)$, $t\in \{1,\cdots,
T\}$, we generate the Gaussian variables ${\bf \hat y}(t)$, $t\in \{1,\cdots,
T\}$.

\item We then estimate the covariance matrix ${\bf \hat \rho}$ of the
Gaussian variables ${\bf \hat y}$, which allows us to compute the variables
$\hat z^2$ and then measure the distance of its estimated distribution
to the $\chi^2$-distribution.

\item Given this covariance matrix ${\bf \hat \rho}$, we generate numerically a
time series of $T$ Gaussian random vectors with the same covariance
matrix $\hat \rho$.

\item For the time series of Gaussian vectors
synthetically generated with covariance matrix ${\bf \hat \rho}$,
we estimate its sample covariance matrix ${\bf \tilde \rho}$.

\item To each of the $T$ vectors of the synthetic Gaussian time series,
we associate the corresponding realization of the random variable
$z^2$, called $\tilde z^2(t)$.

\item We can then construct the empirical distribution for the variable
$\tilde z^2$ and measure the distance between this distribution and
the $\chi^2$-distribution.

\item Repeating 10,000 times the steps 3 to 6,
we obtain an accurate estimate of the cumulative distribution of 
distances between the
distribution of the synthetic Gaussian variables and the theoretical 
$\chi^2$-distribution.

\item Then, the distance obtained at step 2 for the true variables
can be transformed into a significance level by reading the value of 
this synthetically
determined distribution of distances between the distribution of the
synthetic Gaussian variables and the theoretical
$\chi^2$-distribution as a function of
the distance: this provides the probability to observe a distance smaller than
the chosen or empirically determined distance.

\end{enumerate}

\subsection{Sensitivity of the method}
\label{section:sens}

Before presenting the statistical tests, it is important to
investigate the sensitivity of our
testing procedure. More precisely, can we distinguish for instance
between a Gaussian copula
and a Student's copula with a large number of degrees of freedom, for a given
value of the correlation coefficient? Formaly, denoting by $H_\nu$ the
hypothesis according to which the true copula of the data is the Student's
copula with $\nu$ degrees of freedom, we want to determine the minimum
significance level allowing us to distinguish between $H_0$ and $H_\nu$.

\subsubsection{Importance of the distinction between Gaussian and
Student's copulas}

This question has important practical implications because,
as discussed in section \ref{sec:stud_cop}, the Student's copula
presents a significant tail
dependence while the Gaussian copula has no asymptotic tail
dependence. Therefore, if our
tests are unable to distinguish between a
Student's and a Gaussian copula, we may be led to choose the later for the sake
of simplicity and parsimony and, as a consequence, we may
underestimate severely the dependence
between extreme events if the correct description turns out to be the
Student's copula.
This may have catastrophic consequences in risk assessment and
portfolio management.

Figure \ref{fig:1} provides a quantification of the dangers incurred
by mistaking
a Student's copula for a Gaussian one. Consider the case of a
Student's copula with $\nu=20$ degrees of freedom with a correlation 
coefficient
$\rho$ lower than $0.3 \sim 0.4$~; its tail dependence $\lambda_{\nu}(\rho)$
turns out to be less than $0.7\%$, i.e., the probability that one
variable becomes
extreme knowing that the other one is extreme is less than $0.7\%$.
In this case.
the Gaussian copula with zero probability of simultaneous extreme events
is not a bad approximation of the
Student's copula. In contrast, let us take a correlation $\rho$ larger
than $0.7-0.8$ for which the tail dependence becomes larger than
$10\%$, corresponding
to a non-negligible probability of simultaneous extreme events. The effect
of tail dependence becomes of course much stronger as the number
$\nu$ of degrees
of freedom decreases.

These examples stress the importance of knowing
whether our testing procedure allows us to distinguish between a
Student's copula  with $\nu=20$ (or less) degrees of freedom and a
given correlation coefficient
$\rho=0.5$, for instance, and a Gaussian copula with an appropriate correlation
coefficient $\rho'$.

\subsubsection{Statistical test on the distinction between Gaussian
and Student's copulas}

To address this question, we have generated 1,000 pairs of time
series of size $T=1250$, each
pair of random variables following a Student's bivariate distribution
with $\nu$ degrees of
freedom and a correlation coefficient $\rho$ between the two simultaneous
variables of the same pair, while the variables along the time axis
are all independent.
We have then applied the previous
testing procedure to each of the pairs of time series.

Specifically, for each pair of time series,
we construct the marginals distributions and transform the Student's
variables $x_i(k)$ into
their Gaussian counterparts $y_i(k)$ via the transformation
(\ref{nfmlqa}). For each
pair $(y_1(k), y_2(k))$, $k \in \{1,\cdots, T\}$, we estimate its correlation
matrix, then construct the time series
with $T$ realizations of the random variable $z^2(k)$ defined in
(\ref{eq:z}). The set of $T$ variables $z^2$ then allows us to construct
the distribution of $z^2$ (with $N=2$)
and to compare it with the $\chi^2$-distribution with
two degrees of freedom. We then measure the
distances $d_1$, $d_2$, $d_3$ and $d_4$ defined by (\ref{eq:d1}-\ref{eq:d4})
between the distribution of $z^2$ and the $\chi^2$-distribution.
Using the 1,000 pairs of such time series with the same $\nu$ and $\rho$,
we then construct the distribution $D_i(d_i)$, $i\in \{1,2,3,4\}$ of 
each of these distances
$d_i$. Using the
previously determined distribution of distances expected for the
synthetic Gaussian variables, we can translate
each distance $d$ obtained for the Student's vectors into a corresponding
Gaussian probability $p$: $p$ is the probability that pairs
of Gaussian random variables with the correlation coefficient $\rho$
have a distance equal to
or larger than the distance
obtained for the Student's vector time series. A small $p$ corresponds
to a clear distinction between Student's and Gaussian vectors, as it
is improbable
that Gaussian vectors exhibit a distance larger than found for the
Student's vectors.
The ``distribution of probabilities'' $D(p) \equiv D(p(d))$
then assesses how often this ``improbable''
event occurs among the set of 1,000 Student's vectors, i.e.,
attempts to quantify
the rarety of such large deviations. In other words, the
``distribution of probabilities''
$D(p)$ gives the number of Student's vectors that exhibit the value
$p$ for the probability
that Gaussian vectors can have a similar or larger distance.
Then, fixing a confidence level $D^*$, this procedure allows us to
reject or not
the null hypothesis that
the empirical vector of returns is described by a Gaussian copula: this will
occur when the observed $p$ gives a ``distribution of probabilities'' $D(p)$
larger than $D^*$.

\begin{figure}
\begin{center}
\includegraphics[width=15cm]{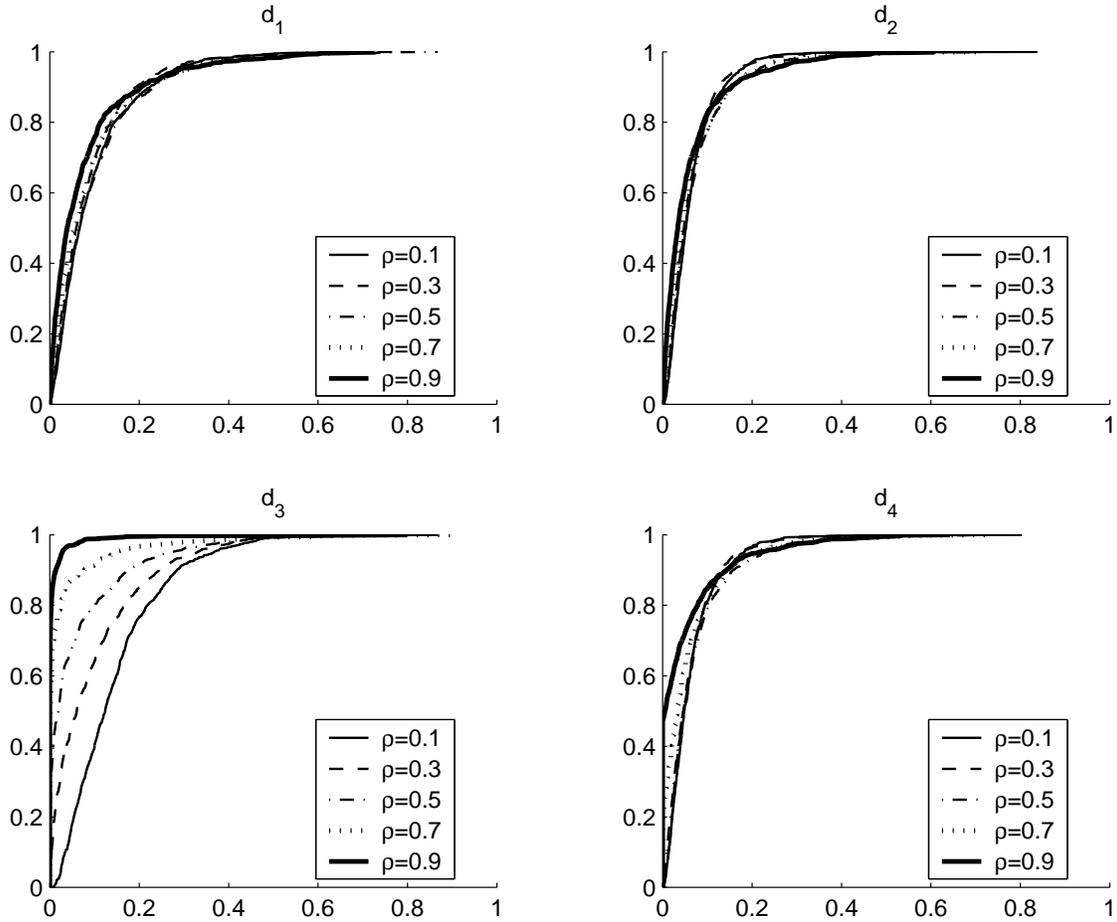}
\caption{\label{fig:ls3} Cumulative ``distribution of probabilities''
$D(p) \equiv D(p(d))$
obtained as the fraction of Student's pairs with $\nu=4$ degrees of freedom
that exhibit the value $p$ for the probability
that Gaussian vectors can have a similar or larger distance. See the text for
a detailled description of how $D(p)$ is defined and constructed.
Each panel corresponds to one of the four distances $d_i$, $i \in
\{1, 2, 3, 4 \}$,
defined in the text by equations (\ref{eq:d1}-\ref{eq:d4}). In each
panel, we construct
the cumulative ``distribution of probabilities'' $D(p)$ for $5$
different values
of the correlation coefficient $\rho=0.1, 0.3,
0.5, 0.7$ and $0.9$ of the Student's copula.}
\end{center}
\end{figure}

\begin{figure}
\begin{center}
\includegraphics[width=15cm]{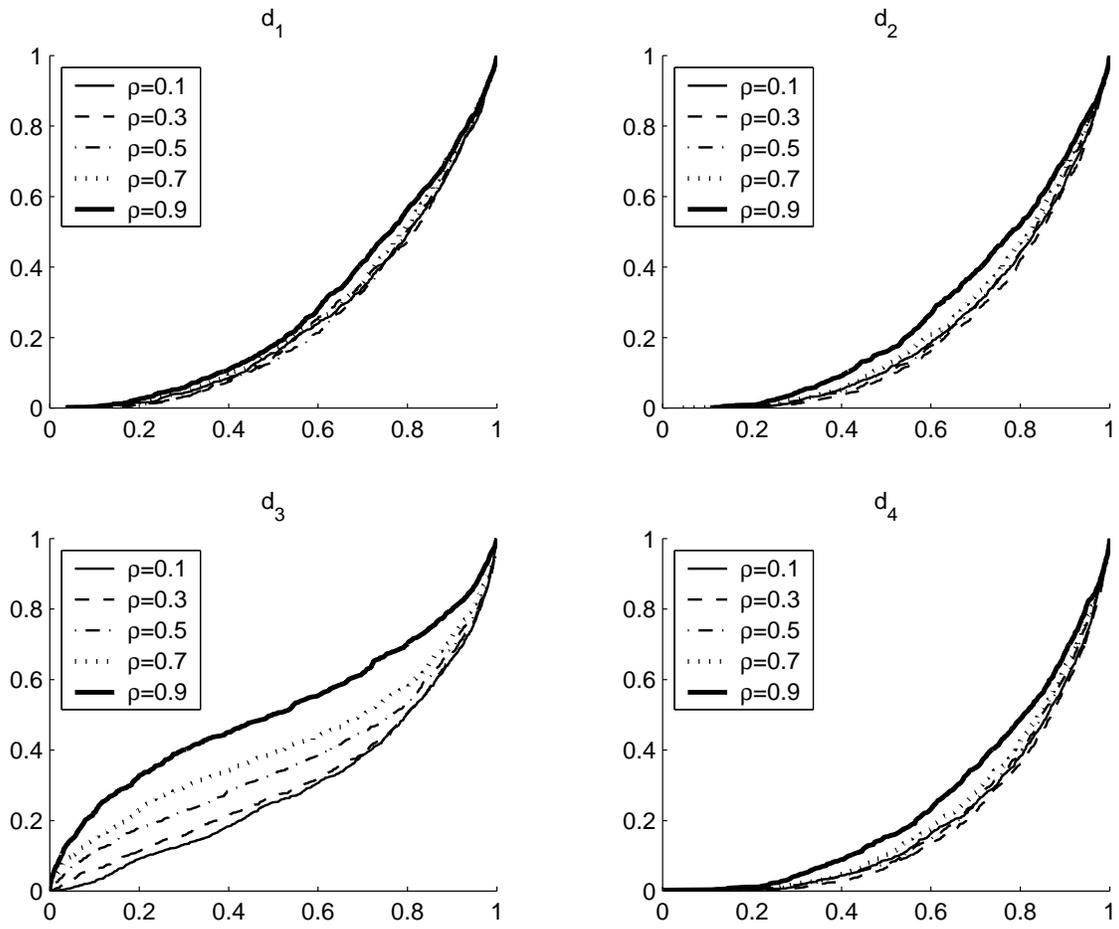}
\caption{\label{fig:ls20} Same as figure \ref{fig:ls3} for
Student's distributions with $\nu=20$ degrees of freedom.}
\end{center}
\end{figure}

The ``distributions of probabilities'' $D(p)$ for each of the four
distances $d_i$, $i\in\{1, 2, 3, 4\}$ are shown
in figure \ref{fig:ls3} for $\nu=4$ degrees of freedom and in
figure \ref{fig:ls20} for $\nu=20$ degrees of freedom, for $5$ different
values of the correlation coefficient $\rho =0.1, 0.3, 0.5, 0.7$ and $0.9$.
The very steep increase observed for almost all cases in figure \ref{fig:ls3}
reflects the fact that most of the 1,000 Student's vectors with
$\nu=4$ degrees of freedom have a small $p$,
i.e., their copula is easily distinguishable from the Gaussian copula. The same
cannot be stated for Student's vectors with $\nu=20$ degrees of freedom.
Note also that the distances $d_1$, $d_2$ and $d_4$ give essentially the
same result while the Anderson-Darling distance $d_3$ is more
sensitive to $\rho$, especially for small $\nu$.

Fixing for instance the confidence level at $D^*=95\%$, we can read from
each of these curves in figures \ref{fig:ls3} and \ref{fig:ls20}
the minimum $p_{95\%}$-value necessary to distinguish a
Student's copula with a given $\nu$ from a Gaussian copula. This $p_{95\%}$
is the abscissa corresponding to the ordinate $D(p_{95\%})=0.95$.
These values $p_{95\%}$ are reported in
table \ref{tab:dist}, for different values of the number $\nu$ of
degrees of freedom ranging from $\nu=3$ to $\nu=50$ and correlation 
coefficients
$\rho=0.1$ to $0.9$. The values of $p_{95\%}(\nu,\rho)$ reported in
table \ref{tab:dist}
are the maximum values that the probability $p$ should take in order
to be able to reject the hypothesis that a Student's copula with $\nu$ degrees
and correlation $\rho$ can be mistaken with a Gaussian copula
at the 95\% confidence level.

\begin{figure}
\begin{center}
\includegraphics[width=15cm]{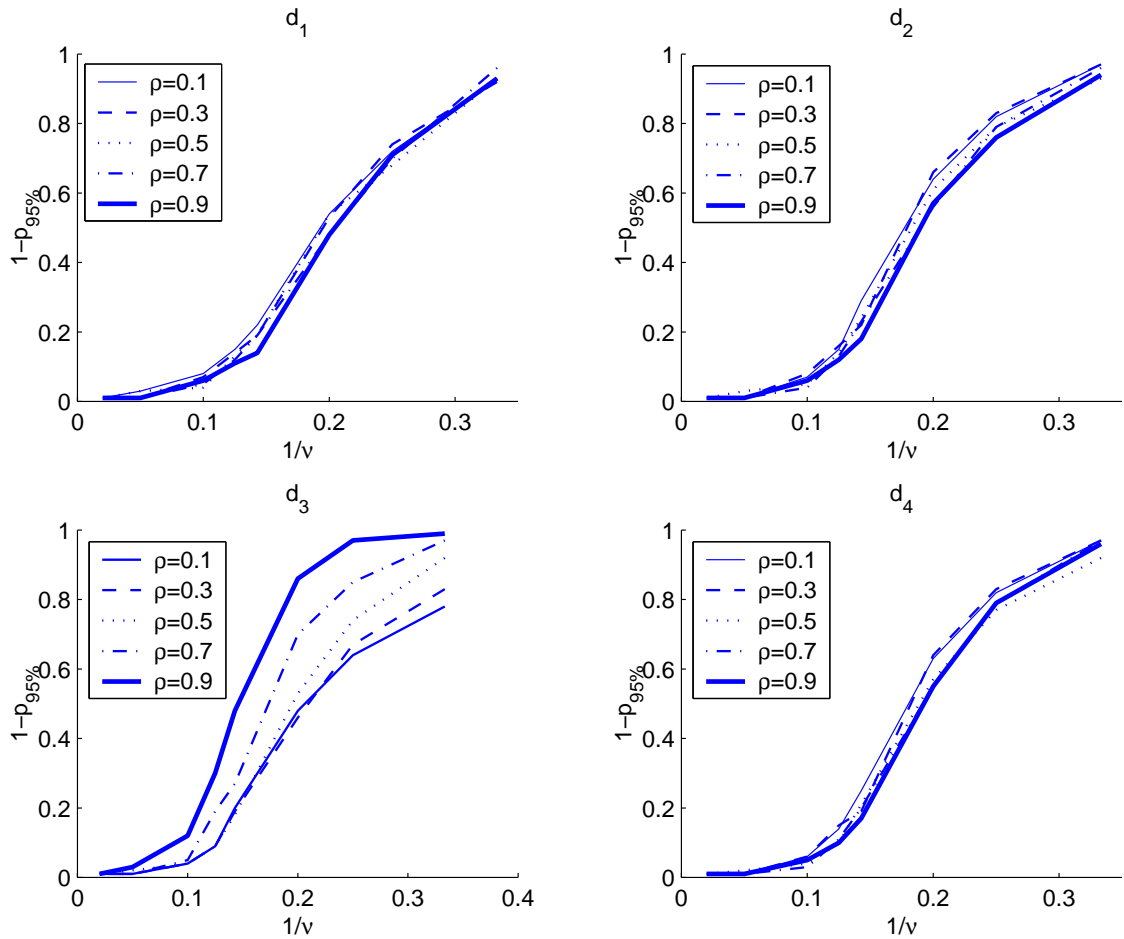}
\caption{\label{fig:ps1} Graph of the minimun significance level ($1-p_{95\%}$)
necessary to distinguish the Gaussian copula hypothesis $H_0$ from the
hypothesis of a student
copula with $\nu$ degrees of freedom, as a function of $1/\nu$, for a given
distance $d_i$ and various correlation coefficients 
$\rho = 0.1, 0.3, 0.5, 0.7$ and $0.9$.} \end{center}
\end{figure}

\begin{figure}
\begin{center}
\includegraphics[width=15cm]{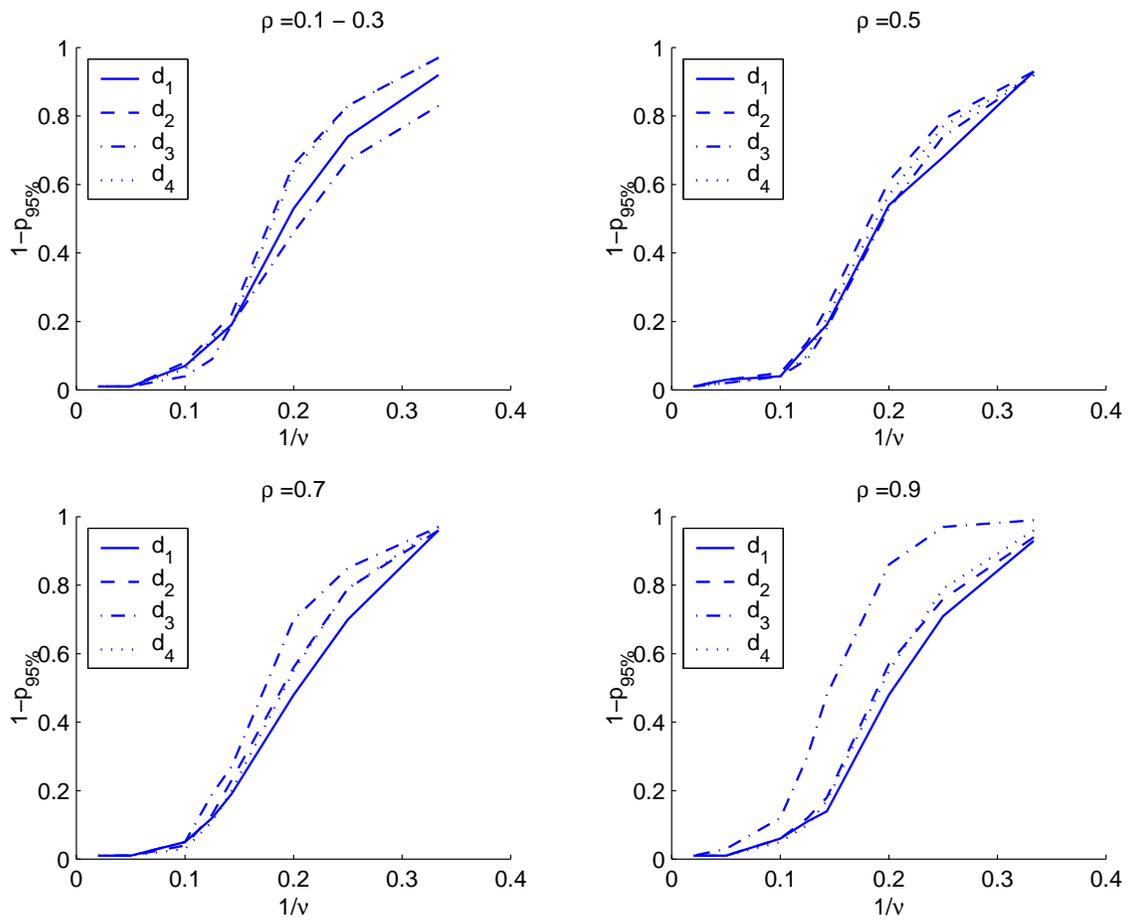}
\caption{\label{fig:ps2}  Same as figure \protect\ref{fig:ps1} but comparing
different distances for the same correlation coefficient $\rho$.}
\end{center}
\end{figure}

The results of the table \ref{tab:dist} are depicted in figures
\ref{fig:ps1}-\ref{fig:ps2} and represent the conventional ``power/size'' statistics.
The statistical ``power'' is usually defined as the rejection of null hypothesis when false.
When the null hypothesis $H_0$ and the alternative hypothesis $H_{\nu}$ are 
identical, the power should be
equal  to $=0.05$, corresponding to the $95\%$ confidence level.
In our framework, this amounts to plot the abscissa as the inverse $\nu^{-1}$
of the number $\nu$ of degrees of freedom, which provides a natural
``distance'' between the Gaussian copula hypothesis $H_0$ and the Student's
copula hypothesis $H_\nu$. In the ordinate, the ``power'' is represented by
the minimum significance level ($1-p_{95\%}$)  necessary to distinguish between
$H_0$ and $H_\nu$. The typical shape of these curves is a sigmoid, starting
from a very small value for $\nu^{-1} \rightarrow 0$, increasing as $\nu^{-1}$
increases and going to 1 as $\nu^{-1}$ becomes large enough. This typical
shape simply expresses the fact that it is easy to separate a Gaussian copula
from a Student's copula with a small number of degrees of freedom, while it is
difficult and even impossible for too large a number of degrees of freedom.

The figure \ref{fig:ps1} shows us that the distances $d_1$, $d_2$ and $d_3$ are
not sensitive to the value of the correlation coefficient $\rho$, while the
discriminating power of $d_3$ increases
with $\rho$. On figure \ref{fig:ps2}, we note that $d_2$ and $d_4$ have the
same discriminating power for all $\rho$'s (which makes them somewhat redundant)
and that they are the most efficient to differentiate $H_\nu$ from $H_0$ for
small $\rho$. When $\rho$ is about 0.5, $d_2$, $d_3$ and
$d_4$ (and maybe $d_1$) are equivalent with respect to the differential power,
while for large $\rho$, $d_3$ becomes the most discriminating one with high
significance.

This study of the test sensitivity involves a non-parametric approach and the
question may arise why it should be prefered to a direct parametric test
involving for instance the calibration of the Student copula. First, a
parametric test of copulas would face the ``curse of dimensionality'', i.e.,
the estimation of functions of several variables. With the limited data set
available, this does not seem a reasonable approach. Second, we have taken the
Student copula as an example of an alternative to the Gaussian copula. However,
our tests are independent of this choice and aim mainly at testing the
rejection of the Gaussian copula hypothesis. They are thus of a more general
nature than would be a parametric test which would be forced to choose one
family of copulas with the problem of excluding others. The parametric test
would then be exposed to the criticism that the rejection of a given choice
might not be of a general nature.

In the sequel, we will choose the level of $95\%$ as the level of
rejection, which leads us to
neglect one extreme event out of twenty. This is not unreasonable in
view of the other significant
sources of errors resulting in particular from
the empirical determination of the marginals and from the presence of
outliers for instance.

\section{Empirical results}

We investigate the following
assets~:
\begin{itemize}
\item foreign exchange rates,
\item metals traded on the London Metal Exchange,
\item stocks traded on the New York Stocks Exchange.
\end{itemize}

\subsection{Currencies}

The sample we have considered is made of  the daily returns for the spot
foreign exchanges for 6 currencies\footnote{The data come from the historical
database of the Federal Reserve Board.}~: the Swiss Franc (CHF), the German
Mark (DEM), the Japanese Yen (JPY), the Malaysian Ringgit (MYR), the Thai Baht
(THA) and the Bristish Pound (UKP). All the exchange rates are expressed
against the US dollar. The time interval runs over ten years, from January 25,
1989 to December 31, 1998, so that each sample contains 2500 data points.

\begin{figure}
\begin{center}
\includegraphics[width=15cm]{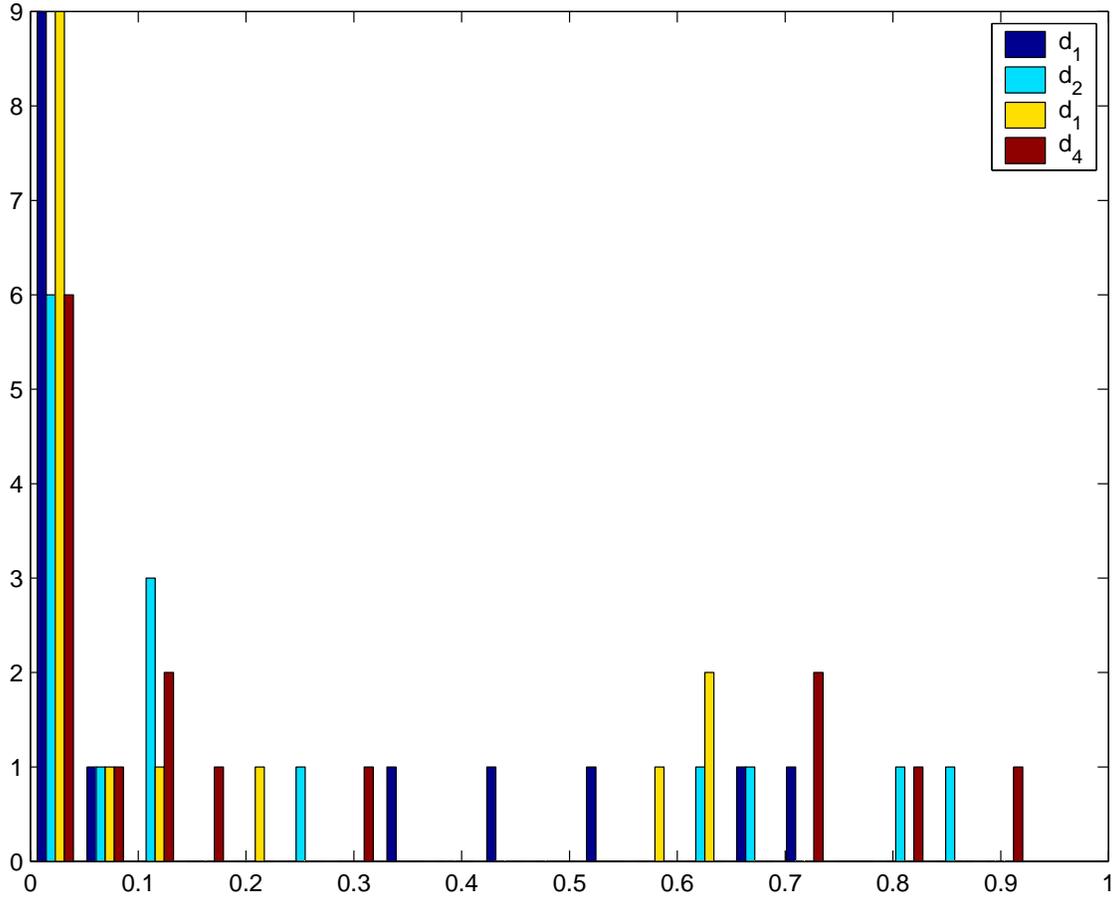}
\caption{\label{fig:curr1} For each distance $d_1$-$d_4$ defined in
equations (\ref{eq:d1})-(\ref{eq:d4}), this figure shows the number
of currency pairs
that give a given $p$ (shown on the abscissa) within a bin interval
of width $0.05$
for different currencies over a 10 years
time interval from January 25, 1989 to December 31, 1998.
$p$ is the probability that pairs
of Gaussian random variables with the same correlation coefficient $\rho$
have a distance (between the distribution of $z^2$ and the theoretical
$chi^2$ distribution) equal to or larger than the corresponding distance
obtained for each currency pair.
A clustering close to the origin signals a significant rejection of
the Gaussian
copula hypothesis.}
\end{center}
\end{figure}

\begin{figure}
\begin{center}
\includegraphics[width=15cm]{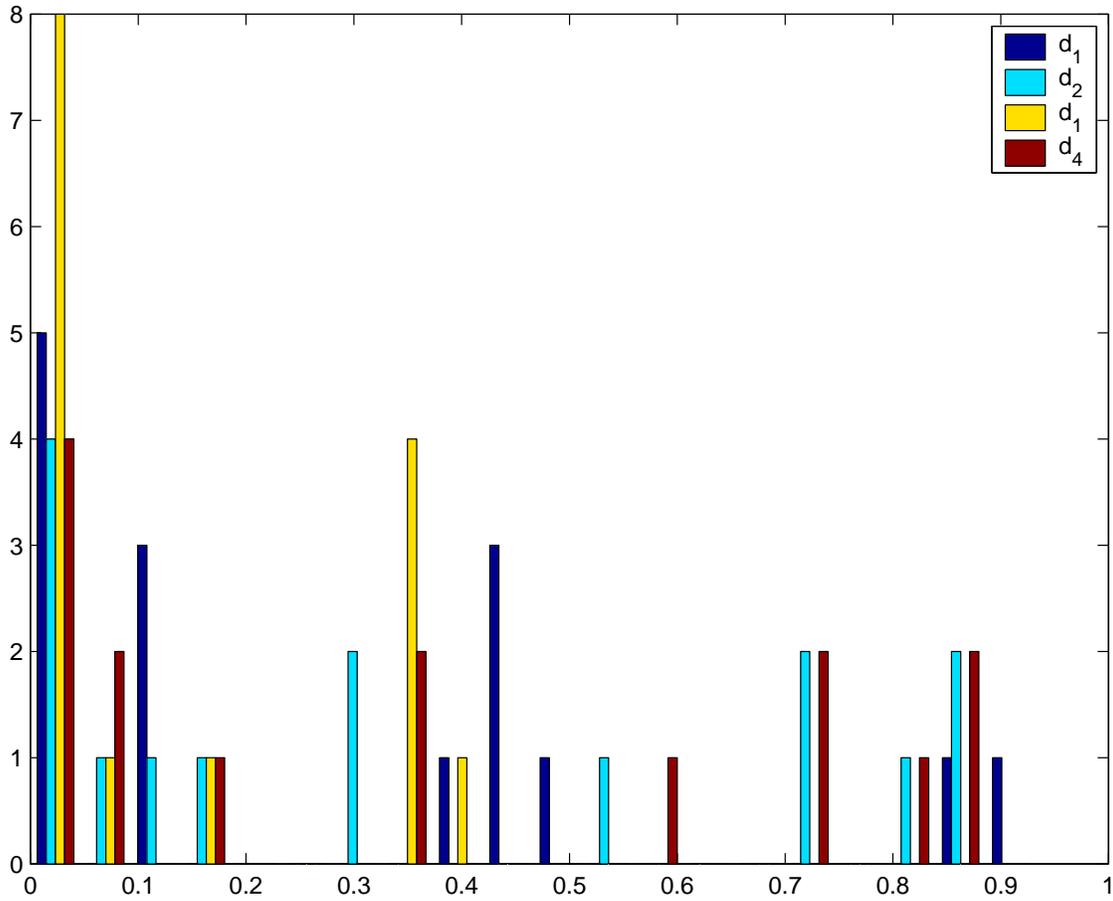}
\caption{\label{fig:curr2} Same as figure \ref{fig:curr1} for
currencies over a 5 years time interval from  January 25, 1989 to January 11,
1994.} \end{center}
\end{figure}

\begin{figure}
\begin{center}
\includegraphics[width=15cm]{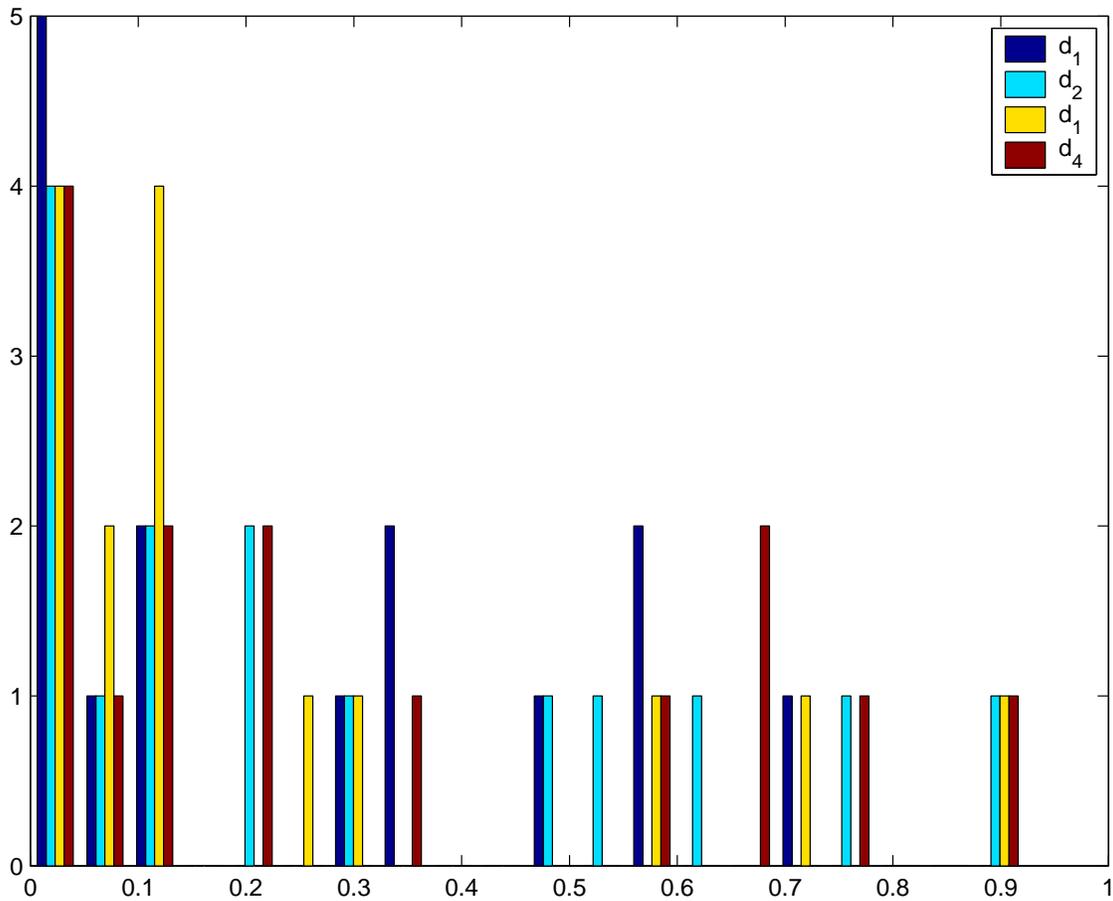}
\caption{\label{fig:curr3} Same as figure \ref{fig:curr1} for
currencies over a 5 years time interval from January 12, 1994 to December
1998.} \end{center}
\end{figure}

We apply our test procedure to the entire sample and to two sub-samples of 1250
data points so that the first one covers the time interval from January 25,
1989 to January 11, 1994 and the second one from January 12, 1994 to December
31, 1998. The results are presented in tables \ref{tab:curr1} to
\ref{tab:curr3} and depicted in figures \ref{fig:curr1} to \ref{fig:curr3}.

Tables \ref{tab:curr1}-\ref{tab:curr3} give, for the total time interval
and for each of the two sub-intervals, the probability $p(d)$ to obtain
from the Gaussian hypothesis a deviation between the distribution of
the $z^2$ and the $\chi^2$-distribution with two degrees of freedom
larger than the observed one for each of the 15 pairs of currencies
according to the distances $d_1$-$d_4$ defined by
(\ref{eq:d1})-(\ref{eq:d4}).

The figures \ref{fig:curr1}-\ref{fig:curr3} organize the information
shown in the tables \ref{tab:curr1}-\ref{tab:curr3} by representing,
for each distance $d_1$ to $d_4$, the
number of currency pairs that give a test-value $p$ within a bin interval
of width $0.05$.
A clustering close to the origin signals a significant rejection of
the Gaussian
copula hypothesis.

At the 95\% significance level, table \ref{tab:curr1} and figure
\ref{fig:curr1} show that only 40\% (according to $d_1$ and $d_3$) but 60\%
(according to $d_2$ and $d_4$) of the tested pairs of currencies
are compatible with the Gaussian copula hypothesis over the entire
time interval.
During the first half-period from January 25, 1989 to Januray 11, 1994 (table
\ref{tab:curr2} and figure \ref{fig:curr2}), 47\% (according to $d_3$) and up
to about 75 \% (according to $d_2$ and $d_4$) of the tested currency pairs
are compatible with the assumption of Gaussian copula, while
during the second sub-period from January 12, 1994 to December 31, 1998 (table
\ref{tab:curr3} and figure
\ref{fig:curr3}),  between 66\% (according to $d_1$) and about 75\%
(according to $d_2$, $d_3$ and $d_4$) of the currency pairs remain compatible
with the Gaussian copula hypothesis. These results raise several
comments both on a statistical and an economic point of view.

We first note that the most significant rejection of the Gaussian copula hypothesis
is obtained for the distance
$d_3$, which is indeed the most sensitive to the events in the tail of the distributions.
The test statistics given by this distance can indeed be very sensitive to the
presence of a single large event in the sample, so much so that
the Gaussian copula hypothesis can be rejected only because of the
presence of this single event (outlier). The
difference between the results given by $d_3$ and $d_4$ (the averaged
$d_3$) are very significant in this respect. Consider for instance the
case of the German Mark and the Swiss Franc. During the time interval
from January 12, 1994 to December 31, 1998, we check on table
\ref{tab:curr3} that the non-rejection probability $p(d)$ is very
significant according to $d_1$, $d_2$ and $d_4$ ($p(d) \ge 31\%$)
while it is very low according to $d_3$: $p(d) = 0.05 \% $, and should
lead to the rejection of the Gaussian copula hypothesis. This
suggests the presence of an outlier in the sample.

\begin{figure}
\begin{center}
\includegraphics[width=15cm]{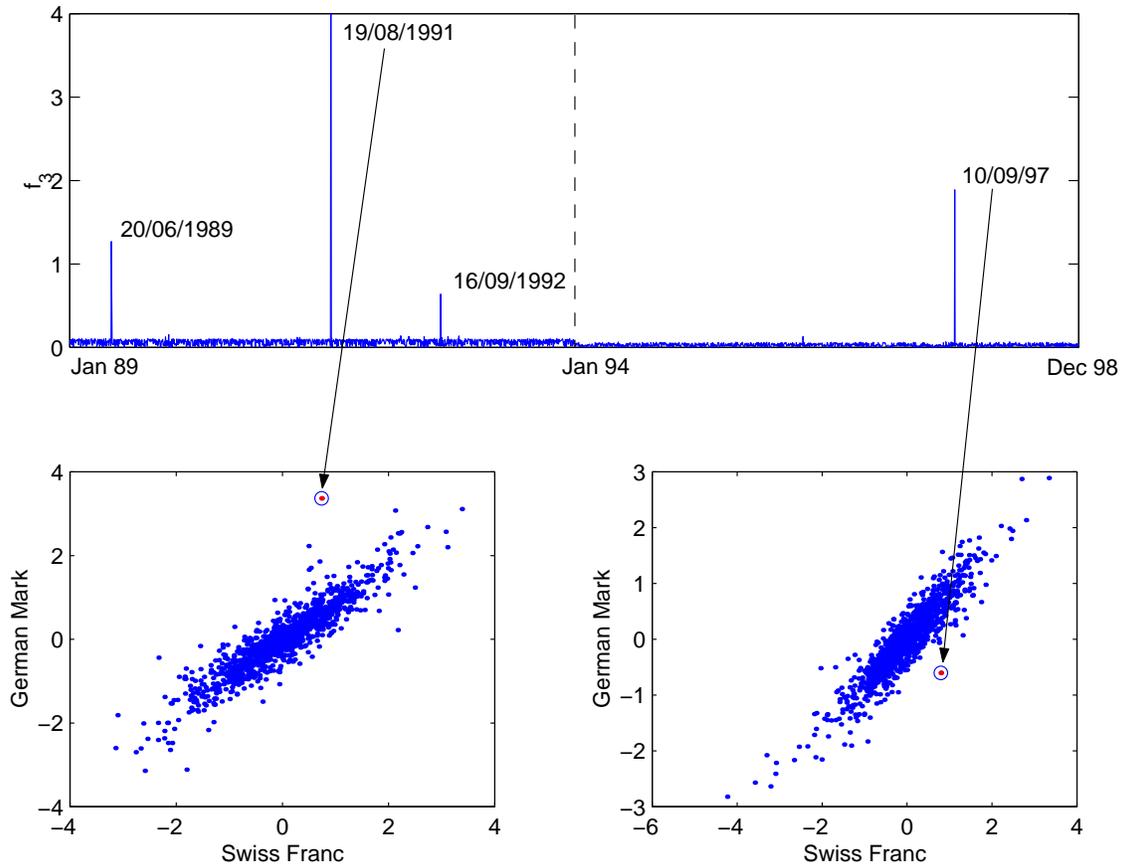}
\caption{\label{fig:dem_chf}  The upper panel represents the graph of the function $f_3(t)$ defined
in (\ref{ghslsl}) used in the definition of the distance $d_3$ for 
the couple Swiss Franc/German Mark as a function of time $t$, over the time intervals from
January 25, 1989 to January 11, 1994 and from January 12, 1994 to
December 31, 1998. The two lower panels represent the scatter plot of
the return of the German Mark versus the return of the Swiss Franc
during the two previous time periods. The circled dot, in each figure,
shows the pair of returns responsible for the largest deviation of
$f_3$ during the considered time interval.}
\end{center}
\end{figure}

To check this hypothesis, we show in the upper panel of  figure \ref{fig:dem_chf} the
function
\be
f_3(t)=\frac{|F_{z^2}(z^2(t)-F_{\chi^2}(\chi^2(t))|}{\sqrt{F_{\chi^2}(\chi^2)[1
- F_{\chi^2}(\chi^2) ] } }~,
\label{ghslsl}
\ee
used in the definition of the Anderson-Darling distance $d_3 =
\max_z f_3(z)$ (see definition (\ref{d3distance})),
expressed in terms of time $t$ rather than $z^2$. The function have
been computed over the two time sub-intervals separately.

 Apart from three extreme
peaks occurring on June $20$, 1989, August $19$, 1991 and September
$16$, 1992 during the first time sub-interval and one extreme peak on
September $10$, 1997 during the second time sub-interval, the
statistical fluctuations measured by $f_3(t)$ remain 
small and of the same order. Excluding
the contribution of these outlier events to $d_3$, the new
statistical significance derived according to $d_3$ becomes similar to 
that obtained with $d_1$, $d_2$ and $d_4$ on each sub-interval. From
the upper pannel of figure \ref{fig:dem_chf}, it is 
clear that the Anderson-Darling distance  $d_3$ is equal to the height 
of the largest
peak corresponding to the event on August $19$, 1991 for the the first
period and to the event on September $10$, 1997 for the second
period. These events are depicted by a circled dot in the two lower
panels of figure \ref{fig:dem_chf}, which represent the return of the
German Mark versus the return of the Swiss Franc over the two
considered time periods.

The event on August $19$, 1991 is  associated with the 
coup against Gorbachev in Moscow: the German mark (respectively
the Swiss franc) lost 3.37\%  (respectively 0.74\%) in daily annualized value
against the US dollar. The 3.37\% drop of the German Mark is the
largest daily move of this currency against the US dollar over the
whole first period.  On September $10$, 1997, the German Mark appreciated by 0.60\%
against the US dollar while the Swiss Franc lost 0.79\% which represents a
moderate move for each currency, but a large joint move. This event is related
to the contradictory announcements of the Swiss National Bank about the
monetary policy, which put an end to a rally of the Swiss Franc along with the
German mark against the US dollar.

Thus, neglecting the large moves associated with major historical
events or events associated with unexpected incoming information, which cannot
be taken into account by a statistical study, we obtain, for $d_3$,
significance levels compatible with those obtained with the other distances.
We can thus conclude that, according to the four distances, during the time interval
from January 12, 1994 to December 31, 1998 the Gaussian copula hypothesis
cannot be rejected for the couple German Mark~/~Swiss Franc.

However, the non-rejection of the Gaussian copula hypothesis does not always
have minor consequences and may even lead to serious problem
in stress scenarios.
As shown in section \ref{section:sens},
the non-rejection of the Gaussian copula hypothesis does not
exclude, at the 95\% significance level, that the
dependence of the currency pairs may be accounted for by
a Student's copula with adequate values of $\nu$ and $\rho$. Still
considering the pair German Mark / Swiss Franc, we see in table
\ref{tab:dist} that, according to $d_1$, $d_2$ and $d_4$, a
Student's copula with about five degrees of freedom allows
to reach the test values given in table \ref{tab:curr3}. But, with the
correlation coefficient 
$\rho=0.92$ for the German Mark/Swiss Franc couple, the Gaussian
copula assumption could lead to neglect a tail dependence coefficient
$\lambda_{5}(0.92)= 63\%$ according to the Student's copula prediction.
Such a large value of $\lambda_{5}(0.92)$ means that when an extreme
event occurs for the German Mark it also occurs for the Swiss Franc with a
probabilty equals to $0.63$. Therefore, a stress scenario based
on a Gaussian copula assumption would fail to account for such coupled
extreme events, which may represent as many as two third of all the
extreme events, if it would turn out that the true copula would be the
Student's copula with five degrees of freedom. In fact, with such a
value of the correlation coefficient, the tail dependence remains high
even if the  number of degrees of fredom reach twenty or more (see
figure \ref{fig:1}).

The case of the Swiss Franc and the
Malaysian Ringgit offers a striking difference. For instance, in the second
half-period, the test statistics $p(d)$ are greater than 70\% and even
reach 91\% while the correlation coefficient is only $\rho=0.16$, so
that a Student's 
copula with 7-10 degrees of freedom can be mistaken with the Gaussian
copula (see table \ref{tab:dist}). Even in the most pessimistic situation
$\nu=7$, the choice of the Gaussian
copula amounts to neglecting a tail dependence coefficient
$\lambda_{5}(0.16)= 4\%$ predicted by the Student's copula.
In this case, stress scenarios based on the Gaussian
copula would predict uncoupled extreme events, which would be shown
wrong only once out of twenty five times.

These two examples show that, more than the number of degrees of
freedom of the Student's copula necessary to describe the data, the
key parameter is the correlation coefficient.

>From an economic point of view, the impact of regulatory mechanisms
between currencies or monetary crisis can be well identified by the rejection
or absence of rejection of our null hypothesis. Indeed, consider the couple German
Mark~/~British Pound. During the first half period, their correlation
coefficient is very high ($\rho=0.82$) and the Gaussian copula hypothesis is
strongly rejected according to the four distances. On the contrary, during the
second half period, the correlation coefficient significantly decreases
($\rho=0.56$) and none of the four distances allows us to reject our null
hypothesis. Such a non-stationarity can be easily explained. Indeed, on
January 1, 1990, the British Pound entered the European Monetary System (EMS),
so that the exchange rate between the German Mark and the Bristish Pound was
not allowed to fluctuate beyond a margin of 2.25\%. However, due to a strong
speculative attack, the British Pound was devaluated on September 1992
and had to leave the EMS.
Thus, between January 1990 and September 1992, the
exchange rate of the German Mark and the British Pound was confined
within a narrow spread, incompatible with the Gaussian copula
description. After 1992, the British Pound exchange rate floated with
respect to German Mark, the dependence between the two
currencies decreased, as shown by their correlation coefficient. In this regime,
we can no more reject the Gaussian copula hypothesis.

The impact of major crisis on the copula can be also clearly identified. Such a
case is exhibited by the couple Malaysian Ringgit/Thai Baht. Indeed, during the
period from Januray 1989 to January 1994, these two currencies have only
undergone moderate and weakly correlated ($\rho=0.29$) fluctuations, so that
our null hypothesis cannot be rejected at the 95\% significance level. On the
contrary, during the period from January 1994 to October 1998, the 
Gaussian copula hypothesis is strongly rejected. This rejection is
obviously due to the persistent and dependent 
($\rho=0.44$) shocks incured by the Asian financial and monetary
markets during the seven months of the Asian Crisis from July 1997 to
January 1998 \cite{FMI,KS99}.

These two cases show that the Gaussian copula hypothesis can be
considered reasonable for currencies in absence of regulatory
mechanisms and of strong and persistent crises.
They also allows us to understand why the results of the test over the entire
sample are so much weaker than the results obtained for the two sub-intervals: 
the time series are strongly non-stationnary.

\subsection{Commodities: metals}

\begin{figure}
\begin{center}
\includegraphics[width=15cm]{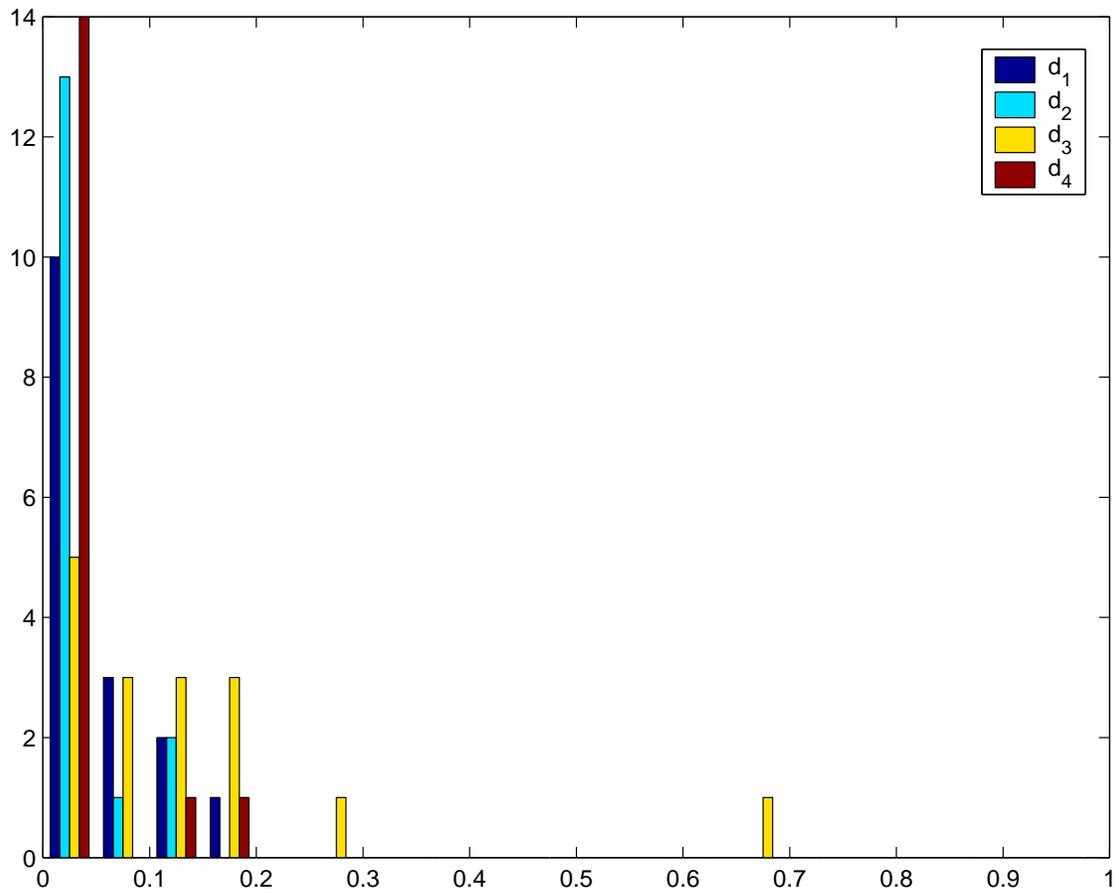}
\caption{\label{fig:lme} Same as figure \ref{fig:curr1} for
metals over a 9 years
time interval from January 4, 1989 to December 30, 1997.}
\end{center}
\end{figure}

We consider a set of 6 metals traded on the London Metal Exchange:
aluminium, copper, lead, nickel, tin and zinc.  Each sample contains 2270 data
points and covers the time interval from January 4, 1989 to December 30, 1997.
The results are synthetized in table \ref{tab:lme} and in figure
\ref{fig:lme}.

Table \ref{tab:lme} gives,
for each of the 15 pairs of commodities,
the probability $p(d)$ to obtain from the Gaussian hypothesis a deviation
between the distribution of the $z^2$ and
the $\chi^2$-distribution with two degrees of freedom larger than the
observed one
for the commodity pair according to the distances $d_1$-$d_4$ defined by
(\ref{eq:d1})-(\ref{eq:d4}).

The figure \ref{fig:lme} organizes the information
shown in table \ref{tab:lme} by representing,
for each distance, the
number of commodity pairs that give a test-value $p$ within a bin interval
of width $0.05$.
A clustering close to the origin signals a significant rejection of
the Gaussian
copula hypothesis.

According to the three distances $d_1, d_2$ and $d_4$, at least two
third and up to $93\%$ of the set of 15 pairs of commodities are
inconsistent with the
Gaussian copula hypothesis. Surprisingly,
according to the distance $d_3$, at the $95\%$ significance level, two
third of the set of 15 pairs of commodities remain compatible with the
Gaussian copula hypothesis.
This is the reverse to the
previous situation found for currencies. These test
values lead to globally reject the Gaussian copula hypothesis.

Moreover, the largest value
obtained for the distance $d_3$ is $p=65\%$ for the pair copper-tin, which
is significantly smaller than the
$80\%$ or $90\%$ reached for some currencies over a similar
time interval. Thus, even in the few cases where the Gaussian copula
assumption is not rejected, the test values obtained are not really
sufficient to distinguish between the Gaussian copula and a Student's
copula with $\nu = 5 \sim 6$ degrees
of freedom. In such a case, with correlation coefficients ranging
between $0.31$ and $0.46$, the tail dependence neglected
by keeping the Gaussian copula is no less than
$10\%$ and can reach $15\%$. One extreme event out
of seven or ten might occur simultaneously on both marginals, which
would be missed by the Gaussian copula.

To summarize, the Gaussian copula does not seem a reasonnable
assumption for metals, and it has not appeared necessary to test these
data over smaller time interval.

\subsection{Stocks}

We now study the daily returns distibutions for 22 stocks among the largest
compagnies quoted on the New York Stock Exchange\footnote{The data come from
the Center for Research in Security Prices (CRSP) database.}: Appl. Materials 
(AMAT), AT\&T (T), Citigroup (C), Coca Cola (KO), EMC, Exxon-Mobil (XOM), Ford
(F), General Electric (GE), General Motors (GM), Hewlett Packard (HPW), IBM,
Intel (INTC), MCI WorldCom (WCOM), Medtronic (MDT), Merck (MRK), Microsoft
(MSFT), Pfizer (PFE), Procter\&Gamble (PG), SBC Communication (SBC), Sun
Microsystem (SUNW), Texas Instruments (TXN), Wal Mart (WMT).

\begin{figure}
\begin{center}
\includegraphics[width=15cm]{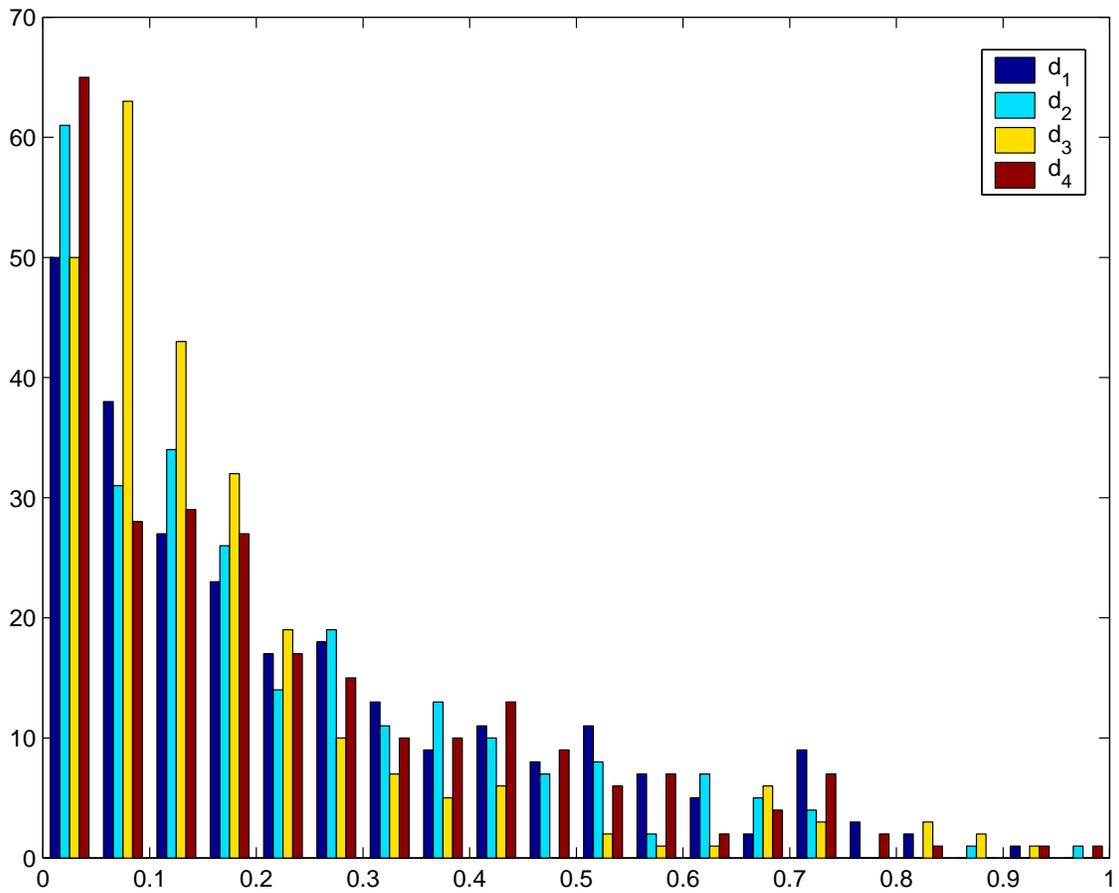}
\caption{\label{fig:stocks1} Same as figure \ref{fig:curr1} for
stocks over a 10 years
time interval from February 8, 1991 to December 29, 2000.} \end{center}
\end{figure}

\begin{figure}
\begin{center}
\includegraphics[width=15cm]{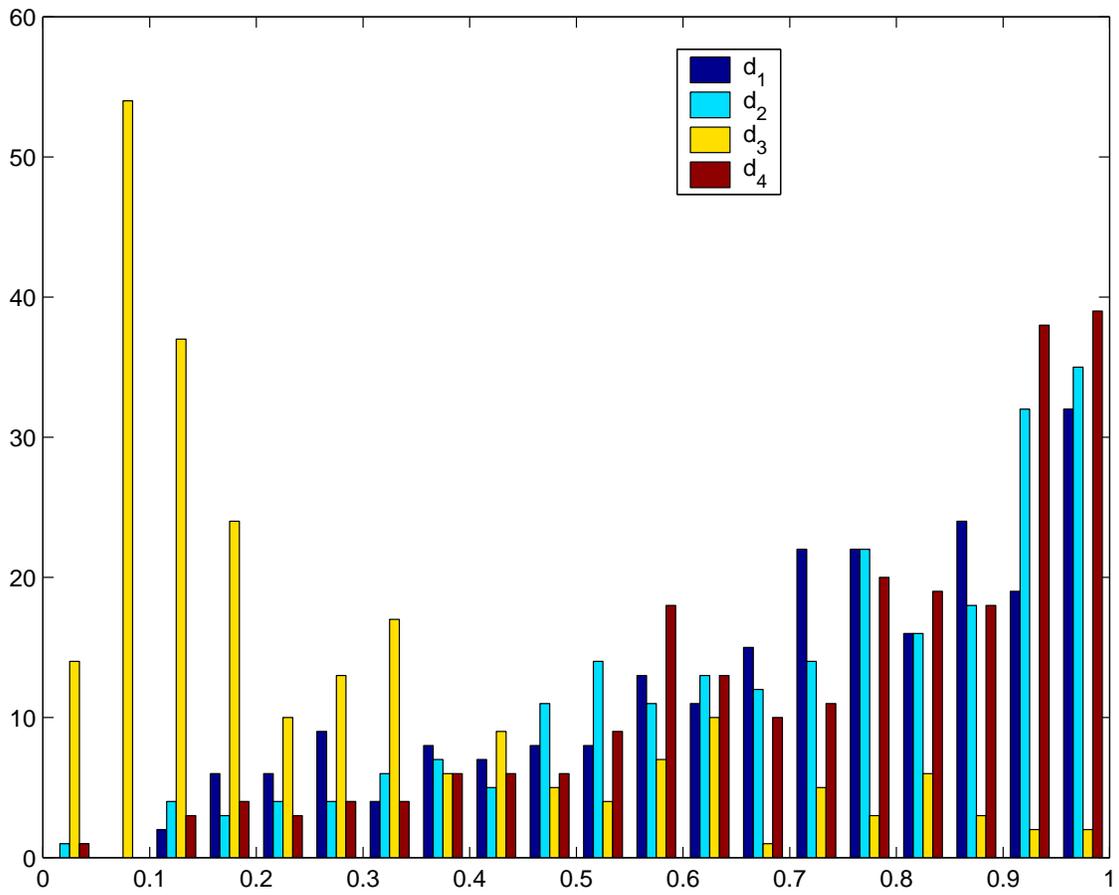}
\caption{\label{fig:stocks2} Same as figure \ref{fig:curr1} for
stocks over a 5 years time interval from February 8, 1991 to January 18, 1996.}
\end{center}
\end{figure}

\begin{figure}
\begin{center}
\includegraphics[width=15cm]{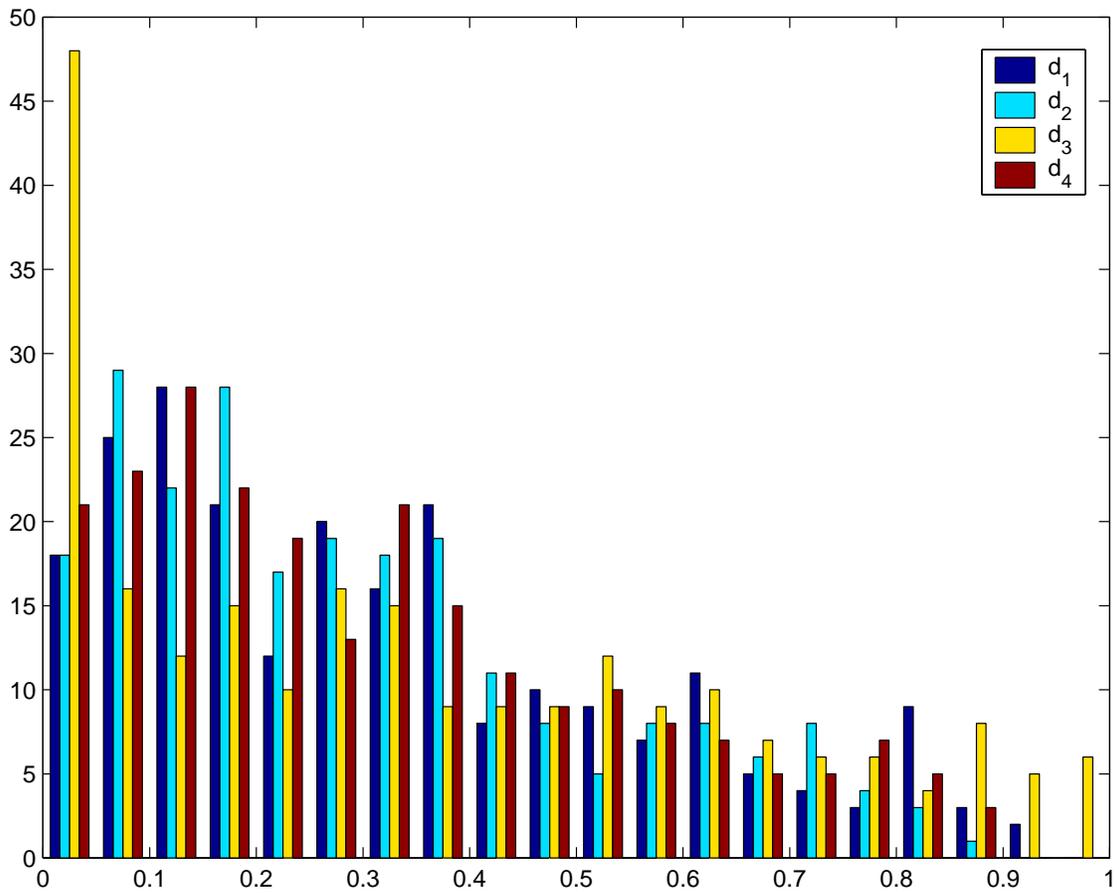}
\caption{\label{fig:stocks3} Same as figure \ref{fig:curr1} for
stocks over a 5 years time interval from January 19, 1996 to December 30,
2000.} \end{center}
\end{figure}

Each sample contains 2500 data points and covers the time interval
from February 8,
1991 to December 29, 2000 and have been divided into two sub-samples of 1250
data points, so that the first one covers the time interval from February 8,
1991 to January 18, 1996 and the second one from January 19, 1996 to December
20, 2000. The results of fifteen randomly chosen pairs of assets are presented
in tables \ref{tab:stock1} to \ref{tab:stock3} while the results obtain
for the entire set are represented in figures \ref{fig:stocks1} to
\ref{fig:stocks3}.

At the 95\% significance level, figure
\ref{fig:stocks1} shows that 75\% of the pairs of stocks are
compatible with the Gaussian copula hypothesis.
Figure \ref{fig:stocks2} shows that over the time
interval from February 1991 to January 1996, this percentage becomes
larger than 99\% for $d_1$, $d_2$ and $d_4$ while it equals 94\%
according to $d_3$. It is striking to note that, during this period,
according to $d_1$, $d_2$ and $d_4$, more than a quarter of the
stocks obtain a test-value $p$ larger than 90\%, so that we can
assert that they are completely inconsistent with the Student's copula
hypothesis
for Student's copulas with less than 10 degrees of freedom.
Among this set of stocks, not a single one has a
correlation coefficient larger than $0.4$, so that a scenario based on
the Gaussian copula hypothesis leads to neglecting a tail dependence of
less than $5\%$ as would be predicted by the Student's copula with
$10$ degrees of freedom. In addition, about
$80\%$ of the pairs of stocks lead to a test-value $p$ larger than
$50\%$ according to
the distances $d_1$, $d_2$ and $d_4$, so that as
much as $80\%$ of the pairs of stocks are incompatible with a
Student's copula with a number of degrees of freedom
less than or equal to $5$. Thus, for correlation
coefficients smaller than $0.3$, the Gaussian copula hypothesis leads
to neglecting a tail dependence less than $10\%$.
For correlation
coefficients smaller than $0.1$ which corresponds to $13\%$ of the total
number of pairs, the Gaussian copula hypothesis leads
to neglecting a tail dependence less than $5\%$.

Figure \ref{fig:stocks3} shows that,
over the time interval from January 1996 to December 2000, $92\%$ of the
pairs of stocks are compatible with the Gaussian copula hypothesis
according to $d_1$,
$d_2$ and $d_4$ and more than $79\%$ according to $d_3$. About a quarter of
the pair of stocks have a test-value $p$ larger than $50\%$ according
to the four
measures and thus are inconsistent with a Student's copula with less
than five degrees of freedom.

For completeness, we present in table \ref{tab:stock4} the results of the tests
performed
for five stocks belonging to the computer area~: Hewlett Packard, IBM, Intel,
Microsoft and Sun Microsystem.  We observe that, during the first half period,
all the pairs of stocks qualify the Gaussian copula Hypothesis
at the 95\% significance level. The results
are rather different for the second half period since about $40\%$ of
the pairs of
stocks reject the Gaussian copula hypothesis according to $d_1$, $d_2$ and
$d_3$. This is probably due to the existence of a few shocks, notably
associated with the crash of the ``new economy'' in March-April 2000.

On the whole, it appears however that there is no systematic rejection
of the Gaussian copula hypothesis for
stocks within the same industrial area, notwithstanding the fact that
one can expect stronger correlations
between such stocks than for currencies for instance.

\section{Conclusion}

We have studied the null hypothesis that
the dependence between pairs of financial assets can be modeled
by the Gaussian copula.

Our test procedure is based on the following simple idea. Assuming
that the copula of two assets $X$ and $Y$ is Gaussian, then the multivariate
distribution of $(X,Y)$ can be mapped into a Gaussian multivariate
distribution,
by a transformation of each marginal into a normal distribution, which
leaves the copula of $X$ and $Y$ unchanged. Testing the Gaussian copula
hypothesis is therefore equivalent to the more standard problem of testing a
two-dimensional multivariate
Gaussian distribution. We have used a bootstrap method to determine and
calibrate the test statistics.
Four different measures of distances between distributions, more or
less sensitive to the
departure in the bulk or in the tail of distributions, have been
proposed to quantify the
probability of rejection of our null hypothesis.

Our tests have been performed over three types of assets: currencies,
commodities (metals) and stocks. In most cases, for currencies and stocks, the
Gaussian copula hypothesis can not be rejected at the 95\% confidence level.
For currencies, according to three of the four
distances at least,
\begin{itemize}
\item 40\% of the pairs of currencies, over a 10 years time interval (due to
non-stationnary data),
\item 67\% of the pairs of currencies, over the first 5
years time interval,
\item 73\% of the pairs of currencies, over the second 5
years time interval,
\end{itemize}
are compatible with the Gaussian copula hypothesis.
For stocks, we have shown that
\begin{itemize}
\item 75\% of the pairs of stocks, over a 10 years time interval,
\item 93\% of the pairs of stocks, over the first 5 years time interval,
\item 92\% of the pairs of stocks, over the second 5 years time interval,
\end{itemize}
are compatible with the Gaussian copula hypothesis.
In constrast, the Gaussian copula hypothesis cannot be considered as
reasonable for metals~: between 66\% and 93\% of the pairs of metals
reject the null
hypothesis at the 95\% confidence level.

Notwithstanding the apparent qualification of the Gaussian copula hypothesis
for most of the currencies and the stocks we have analyzed, we must bear
in mind the fact that a non-Gaussian copula cannot be rejected. In particular,
we have shown that a Student's copula can always be mistaken for a Gaussian
copula if its number of degrees of freedom is sufficiently large. Then,
depending on the correlation coefficient, the Student's copula can predict
a non-negligible tail dependence which is completely missed by the
Gaussian copula assumption.
In other words, the Gaussian copula predicts no tail
dependences and therefore does not account for extreme events that may occur
simultaneously but nevertheless too rarely to modify the test statistics.
To quantify the probability for neglecting such events, we have
investigated the situations when one is unable to distinguish between the
Gaussian and Student's copulas for a given number of degrees of
freedom. Our study leads to the conclusion that it may be very dangerous to
embrace blindly the Gaussian copula hypothesis
when the correlation coefficient between the pair of asset is too high
as the tail dependence neglected by the Gaussian copula can be as
large as $0.6$.
In this respect, the
case of the Swiss Franc and the German Mark is striking. The test
values $p$ obtained are very significant (about $33\%$), so that we cannot
mistake the Gaussian copula for a Student's copula with less than 5-7 degrees
of freedom. However, their correlation coefficient is so high ($\rho=0.9$) that
a Student's copula with, say $\nu=30$ degrees of freedom, still has a large
tail dependence.

This remark shows that it is highly desirable to
develop tests that are specific to the detection of a possible tail dependence
between two time series. This task is very difficult but we hope to report
useful progress in the near future. Another approach is to test for other
non-Gaussian copulas, such as the Student's copula.

\newpage

%FIGURE 1

\clearpage

%FIGURE 2

\clearpage

%FIGURE 3

\clearpage

%FIGURE 4

\clearpage

%TABLE 1
\begin{table}
\begin{center}

\begin{tabular*}{16.3cm}{|c c c c}
\hline
$\nu=3$&
\begin{tabular}{||c |c c c c c|}
$\rho$ &   0.1  &  0.3  &  0.5  &  0.7  &  0.9\\
\hline
d$_1$    &   0.07       & 0.08  &  0.07 &  0.04 &  0.07\\
d$_2$    &   0.03       & 0.03  &  0.07 &  0.04 &  0.06\\
d$_3$    &   0.22       & 0.17  &  0.08 &  0.03 &  0.01\\
d$_4$    &   0.03       & 0.03  &  0.08 &  0.03 &  0.04\\
\end{tabular}&
$\nu=4$&
\begin{tabular} {||c |c c c c c|}
$\rho$ &   0.1  &  0.3  &  0.5  &  0.7  &  0.9\\
\hline
d$_1$   &  0.28&    0.26&    0.32&    0.30&    0.29\\
d$_2$   &  0.18&    0.17&    0.21&    0.21&    0.24\\
d$_3$   &  0.36&    0.33&    0.26&    0.15&    0.03\\
d$_4$   &  0.18&    0.17&    0.23&    0.21&    0.21\\
\end{tabular}\\
\hline
$\nu=5$&
\begin{tabular} {||c |c c c c c|}
$\rho$ &   0.1  &  0.3  &  0.5  &  0.7  &  0.9\\
\hline
d$_1$   &  0.46 &  0.47 &  0.46 &  0.52 &  0.52\\
d$_2$   &  0.36 &  0.34 &  0.39 &  0.44 &  0.43\\
d$_3$   &  0.52 &  0.54 &  0.47 &  0.30 &  0.14\\
d$_4$   &  0.37 &  0.36 &  0.43 &  0.45 &  0.45\\
\end{tabular}&
$\nu=7$&
\begin{tabular} {||c |c c c c c|}
$\rho$ &   0.1  &  0.3  &  0.5  &  0.7  &  0.9\\
\hline
d$_1$   &  0.78&    0.81&    0.81&    0.81&    0.86\\
d$_2$   &  0.71&    0.78&    0.76&    0.77&    0.82\\
d$_3$   &  0.80&    0.81&    0.82&    0.73&    0.52\\
d$_4$   &  0.75&    0.81&    0.79&    0.80&    0.83\\
\end{tabular}\\
\hline
$\nu=8$&
\begin{tabular} {||c |c c c c c|}
$\rho$ &   0.1  &  0.3  &  0.5  &  0.7  &  0.9\\
\hline
d$_1$   &  0.85&    0.86&    0.87&    0.88&    0.89\\
d$_2$   &  0.85&    0.84&    0.86&    0.87&    0.88\\
d$_3$   &  0.91&    0.91&    0.91&    0.81&    0.70\\
d$_4$   &  0.86&    0.85&    0.90&    0.89&    0.90\\
\end{tabular}&
$\nu=10$&
\begin{tabular} {||c |c c c c c|}
$\rho$ &   0.1  &  0.3  &  0.5  &  0.7  &  0.9\\
\hline
d$_1$   &  0.92 &  0.93 &  0.96 &  0.95 &  0.94\\
d$_2$   &  0.93 &  0.92 &  0.95 &  0.96 &  0.94\\
d$_3$   &  0.96 &  0.96 &  0.96 &  0.95 &  0.88\\
d$_4$   &  0.94 &  0.94 &  0.96 &  0.97 &  0.95\\
\end{tabular}\\
\hline
$\nu=20$&
\begin{tabular} {||c |c c c c c|}
$\rho$ &   0.1  &  0.3  &  0.5  &  0.7  &  0.9\\
\hline
d$_1$   &  0.97 &  0.99 &  0.97 &  0.99 &  0.99\\
d$_2$   &  0.99 &  0.99 &  0.97 &  0.99 &  0.99\\
d$_3$   &  0.99 &  0.99 &  0.98 &  0.99 &  0.97\\
d$_4$   &  0.99 &  0.99 &  0.98 &  0.99 & 0.99\\
\end{tabular}&
$\nu=50$&
\begin{tabular} {||c |c c c c c|}
$\rho$ &   0.1  &  0.3  &  0.5  &  0.7  &  0.9\\
\hline
d$_1$   &  0.99&  0.99&  0.99  &  0.99      &  0.99\\
d$_2$   &  0.99&  0.99&  0.99  &  0.99      &  0.99\\
d$_3$   &  0.99&  0.99&  0.99  &  0.99      &  0.99\\
d$_4$   &  0.99&  0.99&  0.99  &  0.99      &  0.99\\
\end{tabular}\\
\hline
\end{tabular*}
\end{center}
\caption{\label{tab:dist} The values $p_{95\%}(\nu,\rho)$ shown in
this table give
the maximum values that the probability $p$ should take in order
to be able to reject the hypothesis that a Student's copula with $\nu$ degrees
and correlation $\rho$ is undistinguishable from a Gaussian copula
at the 95\% confidence level. $p_{95\%}$
is the abscissa corresponding to the ordinate $D(p_{95\%})=0.95$ shown in
figures \ref{fig:ls3} and \ref{fig:ls20}.
$p$ is the probability that pairs
of Gaussian random variables with the correlation coefficient $\rho$
have a distance (between the distribution of $z^2$ and the theoretical $\chi^2$
distribution) equal to
or larger than the corresponding distance
obtained for the Student's vector time series. A small $p$ corresponds
to a clear distinction between Student's and Gaussian vectors, as it
is improbable
that Gaussian vectors exhibit a distance larger than found for the
Student's vectors.
Different values of the number $\nu$ of
degrees of freedom ranging from $\nu=3$ to $\nu=50$ and of the
correlation coefficient
$\rho=0.1$ to $0.9$ are shown. Let us take for instance the example with
$\nu = 4$ and $\rho = 0.3$. The table indicates that $p$ should be less
than about $0.3$ (resp. $0.2$) according to the distances $d_1$ and $d_3$
(resp. $d_2$ and $d_4$) for being able to distinguish this Student's copula
from the Gaussian copula at the 95\% confidence level. This means
that less than $20-30\%$
of Gaussian vectors should have a distance for their $z^2$ larger than the one
found for the Student's. See text for further explanations.
}
\end{table}

\clearpage

%FIGURE 5

\clearpage

%FIGURE 6

\clearpage

%TABLE 2
\begin{table}
\begin{center}
\begin{tabular} {|c c|c c c c c|}
\hline
   &      &       $\hat \rho$ &   d$_1$   &       d$_2$ &  d$_3$ &  d$_4$ \\
\hline
  CHF  & DEM & 0.92 &1.01e-02& 6.70e-03& 0.00e+00 &7.20e-03\\
  CHF  & JPY & 0.53& 3.44e-01& 2.71e-01 &2.32e-02 &2.83e-01\\
  CHF &  MYR & 0.23 &7.27e-01& 8.71e-01& 5.77e-01 &9.26e-01\\
  CHF &  THA & 0.21 &3.08e-02& 9.47e-02 &3.31e-02& 9.52e-02\\
  CHF  & UKP & 0.69 &2.80e-03 &1.80e-03 &6.00e-04 &1.30e-03\\
  DEM &  JPY & 0.54& 2.26e-02 &1.33e-01 &1.00e-01& 1.51e-01\\
  DEM &  MYR & 0.26& 4.25e-01 &6.77e-01& 6.22e-01& 7.35e-01\\
  DEM  & THA & 0.24& 6.53e-02 &1.35e-01& 3.26e-02& 1.32e-01\\
  DEM &  UKP & 0.72 &1.70e-03& 4.00e-04& 0.00e+00 &4.00e-04\\
  JPY &  MYR & 0.31& 2.45e-02& 6.34e-02 &2.26e-01 &6.86e-02\\
  JPY  & THA & 0.34 &0.00e+00 &0.00e+00 &3.24e-02& 0.00e+00\\
  JPY &  UKP & 0.41& 2.85e-02& 3.72e-02 &5.22e-02 &3.09e-02\\
  MYR &  THA & 0.40 &0.00e+00 &0.00e+00& 2.22e-02& 0.00e+00\\
  MYR &  UKP & 0.21& 6.94e-01& 7.94e-01& 6.23e-01& 8.31e-01\\
  THA &  UKP & 0.15& 5.22e-01& 6.23e-01 &3.21e-02 &7.05e-01\\

\hline
\end{tabular}
\end{center}
\caption{\label{tab:curr1} Each row gives the statistics of our test
for each of the 15 pairs of currencies over a 10 years
time interval from January 25, 1989 to December 31, 1998. The column
${\hat \rho}$ gives the empirical correlation coefficient for each pair determined
as in section \protect\ref{sectonrhodef} and defined in (\ref{eq:rho}).
The columns $d_1, d_2, d_3$ and $d_4$
gives the probability to obtain, from the Gaussian hypothesis, a deviation
between the distribution of the $z^2$ and
the $\chi^2$-distribution with two degrees of freedom larger than the
observed one
for the currency pair according to the distances $d_1$-$d_4$ defined by
(\ref{eq:d1})-(\ref{eq:d4}). }
\end{table}

\clearpage

%TABLE 3
\begin{table}
\begin{center}
\begin{tabular} {|c c|c c c c c|}
\hline
   &      &       $\hat \rho$ &   d$_1$   &       d$_2$ &  d$_3$ &  d$_4$ \\
\hline
  CHF  & DEM & 0.92& 1.73e-02& 1.33e-02 &0.00e+00 &1.31e-02\\
  CHF  & JPY & 0.55 &1.34e-01 &1.49e-01& 3.83e-01 &1.41e-01\\
  CHF &  MYR & 0.32& 8.47e-01& 7.00e-01& 3.56e-01 &7.40e-01\\
  CHF  & THA & 0.17 &4.40e-01& 7.10e-01& 3.53e-02 &7.11e-01\\
  CHF  & UKP & 0.79& 3.10e-03 &1.00e-03 &0.00e+00 &5.00e-04\\
  DEM &  JPY & 0.56 &2.46e-02 &9.43e-02 &1.63e-01 &9.26e-02\\
  DEM &  MYR & 0.35& 9.32e-01 &7.95e-01& 3.51e-01& 7.95e-01\\
  DEM &  THA & 0.21 &4.36e-01& 8.77e-01& 3.47e-02& 8.74e-01\\
  DEM &  UKP & 0.82& 0.00e+00 &0.00e+00& 0.00e+00 &0.00e+00\\
  JPY &  MYR & 0.34 &4.90e-01 &5.49e-01 &3.66e-01 &5.94e-01\\
  JPY &  THA & 0.27 &3.89e-01& 3.06e-01& 3.37e-02 &3.59e-01\\
  JPY  & UKP & 0.53 &9.00e-04 &1.66e-02 &6.72e-02 &1.67e-02\\
  MYR &  THA & 0.29 &1.08e-01 &8.71e-02& 3.42e-02& 9.30e-02\\
  MYR &  UKP & 0.33 &1.12e-01& 2.86e-01 &3.54e-01 &3.45e-01\\
  THA &  UKP & 0.21 &4.34e-01 &8.62e-01& 3.13e-02& 8.67e-01\\

\hline
\end{tabular}
\end{center}
\caption{\label{tab:curr2} Same as table \ref{tab:curr1}
for currencies over a 5 years
time interval from January 25, 1989 to Januay 11, 1994. } \end{table}

\clearpage

%TABLE 4
\begin{table}
\begin{center}
\begin{tabular} {|c c|c c c c c|}
\hline
   &      &       $\hat \rho$ &   d$_1$   &       d$_2$ &  d$_3$ &  d$_4$ \\
\hline
CHF &  DEM & 0.92 & 3.15e-01 &3.11e-01& 5.00e-04 &3.41e-01\\
CHF  & JPY &  0.52 & 5.84e-01& 6.44e-01 &1.98e-02 &6.74e-01\\
CHF &  MYR & 0.16  &7.11e-01 &9.15e-01& 8.83e-01& 9.22e-01\\
CHF &  THA & 0.25 &1.10e-02& 3.87e-02& 1.05e-01& 3.34e-02\\
CHF &  UKP & 0.53 & 9.75e-02 &1.03e-01 &2.33e-01 &9.29e-02\\
DEM &  JPY & 0.53 &3.63e-01 &5.40e-01 &1.77e-02 &6.54e-01\\
DEM &  MYR & 0.18 & 3.55e-01 &5.00e-01& 5.84e-01& 5.67e-01\\
DEM & THA & 0.28& 1.28e-02 &2.18e-02 &1.08e-01 &1.51e-02\\
DEM &  UKP & 0.56 &1.15e-01 &1.10e-01 &3.02e-01 &1.06e-01\\
JPY &  MYR & 0.29& 7.63e-02 &2.14e-01 &6.67e-02& 2.23e-01\\
JPY &  THA & 0.38& 0.00e+00 &2.00e-04 &3.09e-02 &2.00e-04\\
JPY &  UKP & 0.28 &4.62e-01& 2.30e-01& 1.23e-01 &2.07e-01\\
MYR &  THA & 0.44 &5.00e-04 &1.20e-03 &5.34e-02 &1.20e-03\\
MYR &  UKP & 0.11 & 5.94e-01& 7.44e-01& 6.95e-01& 7.82e-01\\
THA  & UKP & 0.12 &1.26e-02 &7.66e-02& 1.19e-01 &6.51e-02\\
\hline
\end{tabular}
\end{center}
\caption{\label{tab:curr3} Same as table \ref{tab:curr1}
for currencies over a 5 years
time interval from January 12, 1994 to December 31, 1998.}
\end{table}

\clearpage

%FIGURE 7

\clearpage

%FIGURE 8

\clearpage

%FIGURE 9

\clearpage

%TABLE 5
\begin{table}
\begin{center}
\begin{tabular} {|c c|c c c c c|}
\hline
   &      &       $\hat \rho$ &   d$_1$   &       d$_2$ &  d$_3$ &  d$_4$ \\
\hline
aluminium&       copper&      0.46&   6.46e-02&   4.48e-02&
1.45e-02&   4.00e-02\\
aluminium&       lead     &     0.35&   1.14e-01&   5.01e-02&
1.70e-01&   4.59e-02 \\
aluminium&       nickel   &    0.36&   3.30e-03 &  5.10e-03&
3.41e-02&   6.20e-03  \\
aluminium&       tin        &    0.34&   1.34e-01&   1.38e-01&
1.25e-02&   1.59e-01   \\
aluminium&       zinc     &    0.36&   2.30e-03&   2.20e-03&
6.21e-02&   2.30e-03  \\
copper     &       lead     &   0.35&   4.71e-02&   1.74e-02&
1.79e-01&   1.34e-02 \\
copper     &      nickel    &   0.38&   4.91e-02&   4.60e-02&
1.48e-01&   3.80e-02 \\
copper     &      tin          &   0.32&   1.94e-01&   1.35e-01&
6.53e-01&   1.47e-01\\
copper     &      zinc      &    0.40&   3.24e-02&   2.05e-02&
1.75e-01&   1.94e-02 \\
lead          &     nickel     &   0.32&   6.71e-02&   3.78e-02&
2.74e-01&   3.62e-02 \\
lead          &     tin         &    0.33 &  7.86e-02 &  4.04e-02&
4.91e-02&   3 .31e-02 \\
lead          &     zinc      &    0.42&   2.00e-04&   1.00e-04&
4.59e-02&   3.00e-04 \\
nickel        &    tin         &    0.35&   9.10e-03&   9.20e-03&
8.70e-02&   7.60e-03 \\
nickel        &    zinc      &    0.33&   8.00e-04&   3.40e-03&
8.91e-02&   3.50e-03 \\
tin              &    zinc      &    0.31&   5.30e-03&   2.02e-02&
1.03e-01&   1.75e-02 \\
\hline
\end{tabular}
\end{center}
\caption{\label{tab:lme} Same as table \ref{tab:curr1}
for metals over a 9 years
time interval from January 4, 1989 to December 30, 1997.} \end{table}

\clearpage

%FIGURE 10

\clearpage

%FIGURE 11

\clearpage

%TABLE 6
\begin{table}
\begin{center}
\begin{tabular} {|c c|c c c c c|}
\hline
   &      &       $\hat \rho$ &   d$_1$   &       d$_2$ &  d$_3$ &  d$_4$ \\
\hline
amat    &pfe    &0.15   &7.41e-02       &1.12e-01       &8.40e-03
&1.14e-01\\
c       &sunw   &0.28   &2.56e-01       &4.87e-01       &1.09e-01
&5.39e-01\\
f       &ge     &0.33   &2.52e-01       &2.74e-01       &1.15e-01
&2.90e-01\\
gm      &ibm    &0.21   &1.49e-01       &3.85e-01       &1.62e-01
&4.18e-01\\
hwp     &sbc    &0.12   &4.23e-01       &1.69e-01       &2.52e-01
&1.72e-01\\
intc    &mrk    &0.17   &2.48e-01       &1.09e-01       &6.46e-01
&1.04e-01\\
ko      &sunw   &0.14   &1.41e-01       &1.01e-01       &2.12e-01
&9.35e-02\\
mdt     &t      &0.16   &1.21e-01       &2.81e-01       &8.41e-02
&2.98e-01\\
mrk     &xom    &0.19   &1.54e-01       &1.50e-01       &1.12e-01
&1.45e-01\\
msft    &sunw   &0.44   &3.40e-02       &1.85e-02       &2.60e-03
&1.74e-02\\
pfe     &wmt    &0.27   &4.24e-02       &4.12e-02       &1.54e-01
&3.74e-02\\
t       &wcom   &0.27   &5.67e-02       &8.02e-02       &5.44e-02
&9.07e-02\\
txn     &wcom   &0.28   &4.79e-01       &3.77e-01       &1.52e-01
&3.75e-01\\
wmt     &xom    &0.20   &3.20e-03       &0.00e+00       &6.02e-02
&0.00e+00\\
\hline
\end{tabular}
\end{center}
\caption{\label{tab:stock1} Same as table \ref{tab:curr1}
for stocks over a 10 years
time interval from February 8, 1991 to December 29, 2000. } \end{table}

\clearpage

%TABLE 7
\begin{table}
\begin{center}
\begin{tabular} {|c c|c c c c c|}
\hline
   &      &       $\hat \rho$ &   d$_1$   &       d$_2$ &  d$_3$ &  d$_4$ \\
\hline
amat    &pfe    &       0.10&   5.83e-01&       5.81e-01&
1.18e-01&       6.38e-01\\
c       &sunw   &       0.23&   4.66e-01&       5.94e-01&
4.34e-01&       6.16e-01\\
f       &ge     &       0.31&   8.73e-01&       7.87e-01&
1.54e-01&       8.48e-01\\
gm      &ibm    &       0.21&   6.00e-01&       6.53e-01&
1.03e-01&       5.27e-01\\
hwp     &sbc    &       0.11&   8.73e-01&       8.06e-01&
2.84e-01&       8.59e-01\\
intc    &mrk    &       0.13&   8.59e-01&       8.21e-01&
5.48e-02&       8.65e-01\\
ko      &sunw   &       0.20&   3.53e-01&       5.98e-01&
4.51e-01&       6.79e-01\\
mdt     &t      &       0.14&   9.09e-01&       8.98e-01&
1.68e-01&       9.15e-01\\
mrk     &xom    &       0.12&   5.36e-01&       6.21e-01&
1.20e-01&       6.18e-01\\
msft    &sunw   &       0.40&   2.68e-01&       1.38e-01&
1.60e-01&       1.39e-01\\
pfe     &wmt    &       0.23&   2.94e-01&       4.66e-01&
1.41e-01&       5.23e-01\\
t       &wcom   &       0.19&   7.92e-01&       9.36e-01&
4.95e-02&       9.49e-01\\
txn     &wcom   &       0.23&   9.10e-01&       9.83e-01&
1.00e-01&       9.93e-01\\
wmt     &xom    &       0.22&   7.16e-01&       6.71e-01&
7.35e-02&       6.89e-01\\
\hline
\end{tabular}
\end{center}
\caption{\label{tab:stock2} Same as table \ref{tab:curr1}
for stocks over a 5 years
time interval from February 8, 1991 to January 18, 1996. } \end{table}

\clearpage

%TABLE 8
\begin{table}
\begin{center}
\begin{tabular} {|c c|c c c c c|}
\hline
   &      &       $\hat \rho$ &   d$_1$   &       d$_2$ &  d$_3$ &  d$_4$ \\
\hline
amat    &pfe    &0.19   &2.96e-01       &3.39e-01       &3.10e-02
&3.95e-01\\
c       &sunw   &0.31   &7.12e-01       &6.58e-01       &9.47e-01
&7.08e-01\\
f       &ge     &0.34   &3.80e-01       &2.36e-01       &3.22e-01
&2.18e-01\\
gm      &ibm    &0.21   &3.05e-02       &1.79e-01       &2.37e-01
&2.19e-01\\
hwp     &sbc    &0.11   &3.47e-01       &6.13e-01       &7.17e-01
&6.40e-01\\
intc    &mrk    &0.20   &1.31e-01       &2.06e-01       &5.57e-01
&2.05e-01\\
ko      &sunw   &0.10   &6.89e-01       &3.44e-01       &8.59e-01
&3.52e-01\\
mdt     &t      &0.19   &4.28e-01       &6.11e-01       &5.01e-01
&5.79e-01\\
mrk     &xom    &0.23   &3.57e-01       &6.64e-01       &1.13e-01
&7.38e-01\\
msft    &sunw   &0.46   &5.79e-02       &7.60e-02       &8.00e-04
&8.07e-02\\
pfe     &wmt    &0.30   &2.31e-01       &2.12e-01       &5.59e-01
&1.98e-01\\
t       &wcom   &0.33   &1.20e-01       &1.37e-01       &1.73e-01
&1.40e-01\\
txn     &wcom   &0.31   &5.63e-01       &4.06e-01       &4.64e-01
&4.17e-01\\
wmt     &xom    &0.19   &1.61e-01       &5.38e-02       &3.78e-02
&4.94e-02\\
\hline
\end{tabular}
\end{center}
\caption{\label{tab:stock3} Same as table \ref{tab:curr1}
for stocks over a 5 years
time interval from  January 19, 1996 to December 29, 2000.}
\end{table}

\clearpage

%FIGURE 12

\clearpage

%FIGURE 13

\clearpage

%FIGURE 14

\clearpage

%TABLE 9
\begin{table}
\begin{center}
\begin{tabular*}{14.2cm}{|p{3cm} c}
\hline
Time interval from Frebruary 8, 1991 to January 18, 1996&
\begin{tabular} {|c c|c c c c c|}
   &      &       $\hat \rho$ &   d$_1$   &       d$_2$ &  d$_3$ &  d$_4$ \\
\hline
hwp&        ibm&   0.34&   3.36e-01&   2.26e-01&   3.33e-01&   2.35e-01\\
hwp&       intc&   0.46&   3.01e-01&   4.73e-01&   5.12e-01&   5.21e-01\\
hwp&       msft&   0.41&   7.63e-01&   4.72e-01&   3.23e-01&   4.53e-01\\
hwp&       sunw&   0.40&   2.96e-01&   2.98e-01&   7.66e-01&   3.54e-01\\
ibm&       intc&   0.30&   4.81e-01&   3.54e-01&   4.18e-02&   3.34e-01\\
ibm&       msft&   0.24&   3.93e-01&   6.61e-01&   5.88e-01&   7.07e-01\\
ibm&       sunw&   0.29&   9.65e-01&   9.71e-01&   3.46e-01&   9.86e-01\\
intc&       msft&   0.47&   2.59e-01&   1.45e-01&   4.50e-02&   1.53e-01\\
intc&       sunw&   0.40&   4.81e-01&   3.86e-01&   4.47e-02&   3.95e-01\\
msft&       sunw&   0.40&   2.68e-01&   1.38e-01&   1.66e-01&   1.39e-01\\
\end{tabular}\\
\hline
Time interval from January 19, 1996 to December 29, 2000&
\begin{tabular} {|c c|c c c c c|}
   &      &       $\hat \rho$ &   d$_1$   &       d$_2$ &  d$_3$ &  d$_4$ \\
\hline
hwp&        ibm&   0.46&   2.02e-02&   3.21e-02&   9.60e-03&   3.96e-02\\
hwp&       intc&   0.44&   2.88e-02&   4.89e-02&   6.00e-04&   5.80e-02\\
hwp&       msft&   0.37&   5.23e-02&   9.88e-02&   3.36e-01&   1.18e-01\\
hwp&       sunw&   0.45&   5.66e-01&   5.65e-01&   1.08e-01&   6.23e-01\\
ibm&       intc&   0.43&   5.34e-02&   3.31e-02&   1.68e-02&   2.44e-02\\
ibm&       msft&   0.39&   1.00e-02&   9.50e-03&   2.28e-02&   8.80e-03\\
ibm&       sunw&   0.46&   2.35e-01&   1.56e-01&   3.38e-01&   1.49e-01\\
intc&       msft&   0.57&   3.18e-01&   1.61e-01&   1.15e-01&   1.71e-01\\
intc&       sunw&   0.50&   6.68e-02&   3.55e-02&   1.00e-04&   4.37e-02\\
msft&       sunw&   0.46&   5.79e-02&   7.60e-02&   8.00e-04&   8.07e-02\\
\end{tabular}\\
\hline
\end{tabular*}
\end{center}
\caption{\label{tab:stock4} Same as table \ref{tab:curr1}
for stocks belonging
to the informatic sector, over two time intervals of 5 years.}
\end{table}

\end{document}